\documentclass[preprint,3p]{elsarticle}

\pdfoutput=1

\usepackage{amssymb}
\usepackage{amsmath}

\usepackage{lineno}

\usepackage{appendix}

\usepackage{hyperref}
\usepackage{mathtools}
\usepackage{amsmath,amsfonts,graphicx,amssymb,bm,amsthm}
\usepackage{algorithm,algorithmicx}
\usepackage[noend]{algpseudocode}
\usepackage{fancyhdr}

\usepackage{cases}

\usepackage{subfig} 

\usepackage{hyperref}

\usepackage{makecell}
\usepackage{multirow}

\hypersetup{hidelinks,
	colorlinks=true,
	allcolors=black,
	pdfstartview=Fit,
	breaklinks=true}

\usepackage{indentfirst}
\usepackage{color,xcolor}

\begin{document}

\begin{frontmatter}

\title{Exact Terminal Condition Neural Network for American Option Pricing Based on the Black-Scholes-Merton Equations}
        
\author[1,3]{Wenxuan Zhang\fnref{ce}}
\ead{zhangwenxuan17@mails.ucas.ac.cn}
\author[2,3]{Yixiao Guo\fnref{ce}}
\ead{guoyixiao@lsec.cc.ac.cn}
\author[1,3]{Benzhuo Lu\corref{cor}}
\ead{bzlu@lsec.cc.ac.cn}

\address[1]{SKLMS, ICMSEC, NCMIS, Academy of Mathematics and Systems Science, Chinese Academy of Sciences, Beijing 100190, China}
\address[2]{SKLMS, Institute of Computational Mathematics and Scientific/Engineering Computing, Academy of Mathematics and Systems Science, Chinese Academy of Sciences, Beijing 100190, China}
\address[3]{School of Mathematical Sciences, University of Chinese Academy of Sciences, Beijing 100049, China}
\fntext[ce]{These authors contributed equally.}
\cortext[cor]{Corresponding author}

\begin{abstract}
This paper proposes the Exact Terminal Condition Neural Network (ETCNN), a deep learning framework for accurately pricing American options by solving the Black-Scholes-Merton (BSM) equations. 
The ETCNN incorporates carefully designed functions that ensure the numerical solution not only exactly satisfies the terminal condition of the BSM equations but also matches the non-smooth and singular behavior of the option price near expiration. 
This method effectively addresses the challenges posed by the inequality constraints in the BSM equations and can be easily extended to high-dimensional scenarios. Additionally, input normalization is employed to maintain the homogeneity. Multiple experiments are conducted to demonstrate that the proposed method achieves high accuracy and exhibits robustness across various situations, outperforming both traditional numerical methods and other machine learning approaches.
\end{abstract}


\end{frontmatter}

\section{Introduction}
Options are an important class of financial derivatives, and their valuation is a central issue in quantitative finance. 
Fairly pricing options has been a longstanding challenge, especially for complex options. 
Since Black, Scholes, and Merton proposed the revolutionary Black-Scholes-Merton (BSM) model \cite{black1973pricing, merton1973theory}, the field of option pricing has experienced rapid and significant development.
However, explicit analytical solutions to the BSM equations are available only for a limited number of cases, such as single-asset European options \cite{black1973pricing, merton1973theory}, European options on the maximum or minimum of two assets without dividends \cite{stulz1982options}, European options to exchange one asset for another \cite{margrabe1978value}, and certain European lookback options \cite{conze1991path}. For most other types of options, pricing still relies on numerical and approximate methods.

The pricing of American options is inherently more complicated than that of European options, as option holders have the right to exercise their options at any time prior to expiration.
Explicit solutions for American options are generally far from available.
As a result, numerical approximation techniques, such as the Barone-Adesi-Whaley (BAW) method, 
the binomial tree (BT) model, finite difference (FD) methods, and Monte Carlo (MC) simulations, are widely employed in the industry to price American options.
By dropping a small term in the partial differential equation (PDE), Barone-Adesi and Whaley derived an analytical approximation for American options within the BSM framework \cite{barone1987efficient}.
An early attempt using the binomial tree model was made by Cox, Ross, and Rubinstein (CRR model) \cite{cox1979option}, while \cite{rendleman1979two, boyle1988lattice, trigeorgis1991log, lee2019binomial} proposed some advanced tree methods that converge faster.
Brennan and Schwartz proposed a finite difference method for solving the BSM equations for American put options \cite{brennan1977valuation}. 
Broadie and Glasserman developed a general algorithm based on Monte Carlo simulation to price options \cite{broadie1997pricing}.
Longstaff and Schwartz further refined this approach by introducing the Least Squares Monte Carlo (LSM) method, which prices American options by replacing the future expectation with a least squares interpolation \cite{longstaff2001valuing}.

Compared to single-asset options, multi-asset options are more complicated to price since there are more sources of randomness to consider.
Numerous attempts have been made to employ numerical methods to price them.
The BEG method, introduced by Boyle, Evnine, and Gibbs, utilized the binomial tree method for multi-asset scenarios by incorporating correlated asset price paths across multiple dimensions \cite{boyle1989numerical}.
The finite difference method can also be applied in high-dimensional situations \cite{cole1998generalized}. 
However, both the binomial tree (BT) and finite difference (FD) methods become computationally expensive when extended to high-dimensional cases, especially when more than three underlying assets are involved \cite{moon2008adaptive}. 
Monte Carlo (MC) simulations can be more easily applied in high-dimensional cases, but they suffer from slow convergence rates and face difficulties in accurately addressing the free boundary issue inherent in American options \cite{broadie1997pricing}.

To address these challenges, deep learning approaches have emerged as promising alternatives in recent years.
Dhiman and Hu applied physics-informed neural network (PINN) \cite{RAISSI2019686} to solve BSM equations for single-asset options \cite{dhiman2023physics}, while Gatta et al. extended PINN to multi-asset American put options \cite{gatta2023meshless}.
Sirignano and Spiliopoulos explored the multi-asset BSM equations using the deep Galerkin method (DGM) with similar loss functions \cite{sirignano2018dgm}. 
The deep parametric PDE method \cite{glau2022deep}, developed by Glau and Wunderlich, trains neural networks to approximate BSM solutions across a wide range of input parameters. 
Another influential framework, the deep BSDE method, developed by Han, Jentzen, and E \cite{han2018solving}, reformulates the PDE as a backward stochastic differential equation.
This method has been extended to multi-asset American options by Negyesi and Oosterlee \cite{negyesi2025deep}.
However, the deep BSDE framework is limited to producing solutions at individual points, rather than pricing the entire surface.

In this paper, we focus on  American options and provide approximate solutions over the spatial-temporal domain. 
The problem is formulated as a partial differential equation (PDE) in a linear complementarity form, which includes both equalities and inequalities, as well as a terminal boundary condition. 
Two primary challenges arise when solving these equations. 
First, the presence of inequalities complicates the solution process, as the problem can be reformulated as a Stefan-type free-boundary problem, where the boundaries are unknown \cite{van1976optimal}. 
Second, singular behaviors may arise near maturity due to the non-differentiability of the payoff function.
These singularities cause the derivative of the option price to approach infinity near the terminal, posing challenges for directly using neural network methods to obtain high-accuracy solutions.
As shown in our experiments in Section~\ref{sec4}, the accuracy of the solution deteriorates significantly near these critical points if singularities are not appropriately addressed.

To address these challenges, we introduce the Exact Terminal Condition Neural Network (ETCNN), a deep learning method designed to automatically satisfy the terminal condition. 
The primary approach involves constructing a function $g_2$ that exactly satisfies the terminal condition. The solution is then formulated as the sum of $g_2$ and the product of a certain function $g_1$ and a neural network approximation, where $g_1$ vanishes at the terminal. 
However, our experiments indicate that not all functions satisfying the terminal conditions necessarily yield higher accuracy. The effectiveness of the method is heavily dependent on the smoothness and structure of the residual difference between the true solution and the chosen function $g_2$. 
Therefore, $g_2$ is carefully constructed not only to satisfy the terminal condition, but also to preserve the differentiability characteristics of the exact solution and exhibit appropriate asymptotic behavior as time approaches expiration. 
We provide both mathematical analysis and financial interpretations for designing suitable forms of $g_2$ applicable to a wide range of American option pricing scenarios. 
Additionally, we design an input normalization layer to normalize the underlying asset prices in the input vector of the network, which can further improve the accuracy of the model.

Our approach has the following primary advantages. First, compared to PINN, ETCNN fully aligns with the true solution at the terminal, eliminating boundary condition errors associated with collocation methods and significantly improving solution accuracy. 
Second, our network eliminates the need to include the boundary term in the objective loss function. This removes the hyperparameter assigned to the weight of the boundary loss term, thus simplifying hyperparameter tuning and reducing training complexity.
Third, the method effectively addresses the free-boundary challenge inherent in BSM equations for American options. Rather than directly solving for the free boundary, it enables an accurate determination of the boundary in an indirect way.
Fourth, ETCNN accurately captures the singularity behavior near the terminal, enhancing solution precision in these critical regions.
Finally, our method can be easily extended to high-dimensional cases, maintaining high accuracy in complex multi-asset scenarios where traditional methods are often ineffective. 

We conduct extensive experiments on both single-asset and multi-asset American options, corresponding to low-dimensional and high-dimensional BSM equations. 
These experiments consider various terminal boundary conditions and a wide range of equation coefficients, highlighting the universality and robustness of our method. 
The results demonstrate that our ETCNN improves accuracy by 1-2 orders of magnitude compared to PINN and achieves or even exceeds the accuracy of traditional numerical methods in low-dimensional cases. Moreover, in high-dimensional scenarios where traditional methods face significant challenges, our approach consistently achieves high accuracy.

The structure of this paper is organized as follows. 
Section~\ref{sec2} introduces the mathematical formulation of the BSM model for option pricing and provides an overview of the PINN framework. 
Section~\ref{sec3} presents the architecture of the ETCNN, key criteria for designing the exact terminal function $g_2$, and the methodological formulation for American options.
Section~\ref{sec4} reports the result of numerical experiments, and compares the performance of our method with PINN and other numerical approaches.
Finally, Section~\ref{sec5} summarizes our work and outlines potential directions for future research.

\section{Preliminaries}
\label{sec2}

\subsection{BSM Equations for Option Pricing}
This section presents the BSM equations for option pricing problems. Section~\ref{sec2.1.1} introduces the BSM equations for European options, which have the simplest form of these equations. Section~\ref{sec2.1.2} to \ref{sec2.1.4} focuses on the BSM equations for American options, which are the central theme of this article. These sections will cover low and high-dimensional forms, along with a discussion of their properties.

\subsubsection{BSM Equations for European Options}
\label{sec2.1.1}

This model includes two types of assets, risk-free assets and risky assets. The value of risk-free assets can be described by a deterministic process, 
\begin{equation*}
    dR(t) = rR(t)dt,
\end{equation*}
where $r$ is the risk-free rate and is assumed to be a constant under the BSM model. Risky assets typically serve as the underlying assets for options contracts and include different financial products such as stocks, stock indices, and futures. The BSM model assumes that
the value of risky assets $s = S(t)$ follows a geometric Brownian motion,
\begin{equation*}
    dS(t) = \mu S(t)dt + \sigma S(t)dW(t),
\end{equation*}
where $\mu$ is the drift, $\sigma > 0$ is the volatility, and $W(t)$ is a standard Brownian motion. Both $\sigma$ and $r$ are expressed in annualized terms. In the geometric Brownian motion framework, the drift and volatility are assumed to remain constant.

Let $V = V(s, t) = V(S(t), t)$ be the price of a single-asset European option. According to It$\hat{\text{o}}$'s lemma, 
\begin{equation*}
dV(S(t), t) = \Big(\frac{\partial V}{\partial t} + \mu S(t)\frac{\partial V}{\partial s} + \frac{1}{2}\sigma^2 S(t)^2\frac{\partial^2 V}{\partial s^2}\Big)dt + \sigma S(t)\frac{\partial V}{\partial s}dW(t). 
\end{equation*}
To determine the price of an option with an underlying asset following the above geometric Brownian motion, a self-financing trading strategy is used. This strategy dynamically adjusts the risk-free asset and the underlying asset to replicate the option’s payoff profile. Let $X(t)$ denote the value of a portfolio at time $t$, consisting of $\Delta(t)$ shares of the underlying asset, with the remainder $X(t) - \Delta(t)S(t)$ invested in a risk-free asset at rate $r$.
The self-financing assumption implies that changes in the
portfolio are solely attributed to gains or losses in the underlying securities, with no impact from changes in the holdings \cite{harrison1981martingales}. The 
change of the portfolio is 
\begin{align*}
dX(t) &= \Delta(t)dS(t) + r(X(t)-\Delta(t)S(t))dt \notag \\
&= \big(rX(t) + (\mu - r)\Delta(t)S(t)\big)dt + \sigma \Delta(t)S(t)dW(t).  
\end{align*}

By selecting the hedge ratio $\Delta(t)$ and portfolio value $X(t)$ such that the  terminal payoff of $X(T)$ matches the payoff of the option $V(T)$ and ensuring 
$dX(t) = dV(t)$ for all $t$, the portfolio value $X(t)$ at any time $t$  equals the option's theoretical price. From this, we have the following equations,
\begin{equation*}
\left\{  
\begin{aligned} 
&\quad \frac{\partial V}{\partial t} + \mu S(t)\frac{\partial V}{\partial s} + \frac{1}{2} \sigma^2 S(t)^2 \frac{\partial^2 V}{\partial s^2} = rX(t) + (\mu - r)\Delta(t)S(t), \\ 
&\quad \sigma S(t)\frac{\partial V}{\partial s} = \sigma \Delta(t)S(t), \\ &\quad X(t) = V(t).
\end{aligned}
\right. 
\end{equation*}
By simplifying the above equations and considering the boundary conditions, the celebrated BSM equation for single-asset European options is obtained,
\begin{equation}
\left\{  
\begin{aligned} &\quad
   \frac{\partial V}{\partial t} + \frac{1}{2}\sigma^2s^2 \frac{\partial^2 V}{\partial s^2} + rs\frac{\partial V}{\partial s} - rV = 0, \quad \forall s \geq 0, t \in [0, T),  \\ &\quad  V(s, T) = \Phi(s), \quad \forall s \geq 0.
   \end{aligned}   \label{9}
\right. 
\end{equation}
The solution to this equation yields the price of European options. The boundary condition is applied at the expiration date $t=T$, thus it is referred to as a terminal condition. $\Phi(s)$ is the payoff of the function at $t=T$. Denote\begin{equation*}
    x^+ = \max (x, 0) = \begin{cases}
x&\text{if}\ x \geq 0,\\
0&\text{if}\ x < 0,
\end{cases}  
\end{equation*} as an abbreviation.
For European call options, $\Phi(s) = (s-K)^+$ and for European put options, $\Phi(s) = (K-s)^+$. Here, $K$ is the strike price, which is predetermined in the contract. 

If the underlying asset pays a constant continuous dividend yield $q$, the BSM equation is modified as:
\begin{equation*}
\left\{  
\begin{aligned} &\quad
   \frac{\partial V}{\partial t} + \frac{1}{2}\sigma^2s^2 \frac{\partial^2 V}{\partial s^2} + (r-q)s\frac{\partial V}{\partial s} - rV = 0 , \quad \forall s \geq 0, t \in [0, T), \\ &\quad  V(s, T) = \Phi(s),  \quad \forall s \geq 0.
   \end{aligned}    
\right.  
\end{equation*}
Note that $q$ only appears in the coefficients of the first-order derivative, but not in the coefficients of $V$. Therefore, it is not simply a replacement of $r$ with $r-q$.

\subsubsection{BSM Equations for American Options}
\label{sec2.1.2}

American options differ from European ones in that they can be exercised early. The BSM equations for American options can be formulated similarly to Eq.~(\ref{9}), but with modifications to incorporate their early exercise feature. We define the following operator for the primary equation,
\begin{equation}   \mathcal{F}\big(V(s,t)\big) = \frac{\partial V}{\partial t} + \frac{1}{2}\sigma^2 s^2 \frac{\partial^2 V}{\partial s^2} + (r-q)s\frac{\partial V}{\partial s}-rV.  \label{9'}
\end{equation}
The terminal condition $\Phi(s)$ also represents the intrinsic value, as it is the payoff obtained by exercising the option immediately. The time value of the option is defined as the difference between an option price and its intrinsic value, expressed as follows,  
\begin{equation*}
   \mathcal{TV}\big(V(s, t)\big) = V(s, t) - \Phi(s). 
\end{equation*}
The time value of an American option must always be nonnegative to avoid risk-free arbitrage opportunities. Otherwise, arbitrageurs can buy options at price $V(s,t)$, exercise them immediately to receive the payoff $\Phi(s)$, and secure a risk-free positive profit of $\Phi(s) -V(s,t)$. To prevent arbitrage opportunities, the price of American options $V(s, t)$ satisfies the following linear complementarity conditions 
\cite{shreve2004stochastic}, \begin{subequations}
\begin{numcases}{}
\quad \mathcal{F}\big(V(s,t)\big) \leq 0, \quad \forall s \geq 0, t \in [0, T), \label{17} \\ 
\quad   \mathcal{TV}\big(V(s, t)\big) \geq 0,\quad \forall s \geq 0,  t \in [0, T), \label{16} \\ \quad 
\mathcal{F}\big(V(s,t)\big)\cdot \mathcal{TV}\big(V(s, t)\big) = 0,\quad \forall s \geq 0, t \in [0, T),  \label{11'}\\
\quad V(s, T) = \Phi(s), \quad \forall s \geq 0 \label{18'},
\end{numcases}
\end{subequations}
These complementarity conditions introduce additional complexities compared to Eq.~(\ref{9}). Specifically, they contain inequality constraints that complicate both analytical and numerical methods. Moreover, Eq.~(\ref{11'}) ensures that at least one of the inequalities must hold with equality, but the specific regions where each inequality becomes an equality are unknown in advance. Additionally, $\Phi(s)$ is typically non-differentiable, which adds further challenges in numerical approximation.

\subsubsection{Free Boundary Property in American Options}
\label{sec2.1.3}

The inequality constraints give rise to the free boundary problem, which poses a significant challenge in the valuation of American options. To illustrate this, consider the case of an American put option. When the price of the underlying asset $s$ is low enough, immediate exercise of the option yields a payoff that exceeds the expected benefit of holding the option. This region is referred to as the stopping region. 
There exists a function $S^*(t)$, 
which defines the stopping region as the set of points below its curve, \begin{equation*}
    \mathcal{S} = \{0 \leq s \leq S^*(t), 0 \leq t \leq T\} = \{(s,t):V(s,t) = (K-s)^+\}.  
\end{equation*}
The complementary region is called the continuation region, denoted by
\begin{equation*}
  \mathcal{C} = \{s > S^*(t), 0 \leq t \leq T\} = \{(s,t):V(s,t) > (K-s)^+\}. 
\end{equation*}
However, $S^*(t)$ is an unknown function, thus, its function curve is called a free boundary.

In the stopping region, Eq.~(\ref{16}) holds equal. 
If $(s,t)$ falls into this area, the optimal strategy for the option holder is to exercise the option immediately. Hence, $S^*(t)$ is also referred to as the optimal exercise boundary \cite{gatta2023meshless}. Conversely, in the continuation region, Eq.~(\ref{17}) holds equal and $V(s, t) > (K-s)^+$.
In this region, the holder will choose to hold the option. $S^*(t)$ represents the maximum price at which the holder will choose to exercise the option early, i.e. 
\begin{equation}
    S^*(t) = \sup \{s\in \mathbb{R}^+: V(s, t) = (K-s)^+ \}. \label{15}
\end{equation}

\begin{figure}[!ht]
  \subfloat[]{\includegraphics[width=0.47\textwidth]{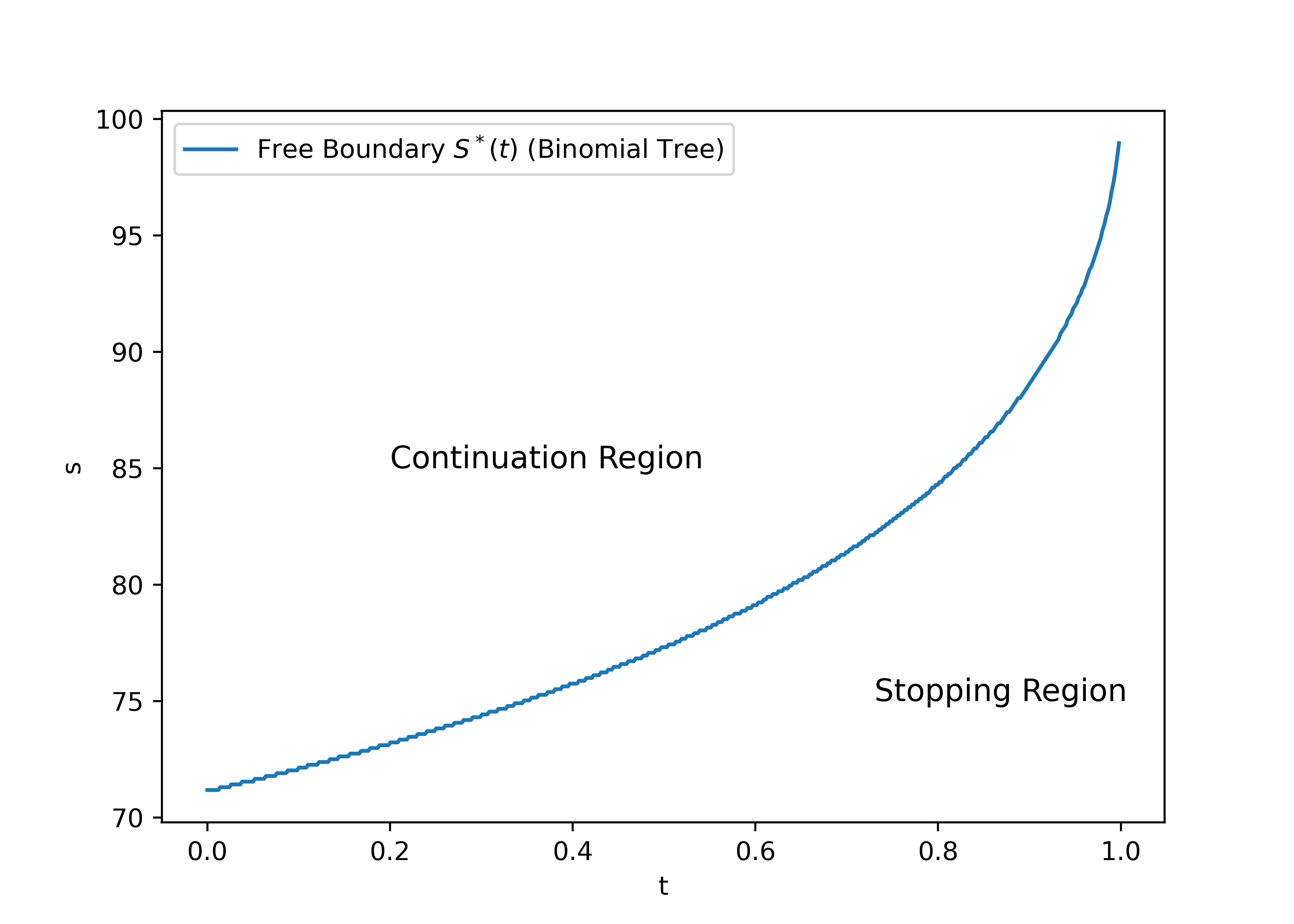}}
 \hfill 	
  \subfloat[]{\includegraphics[width=0.47\textwidth]{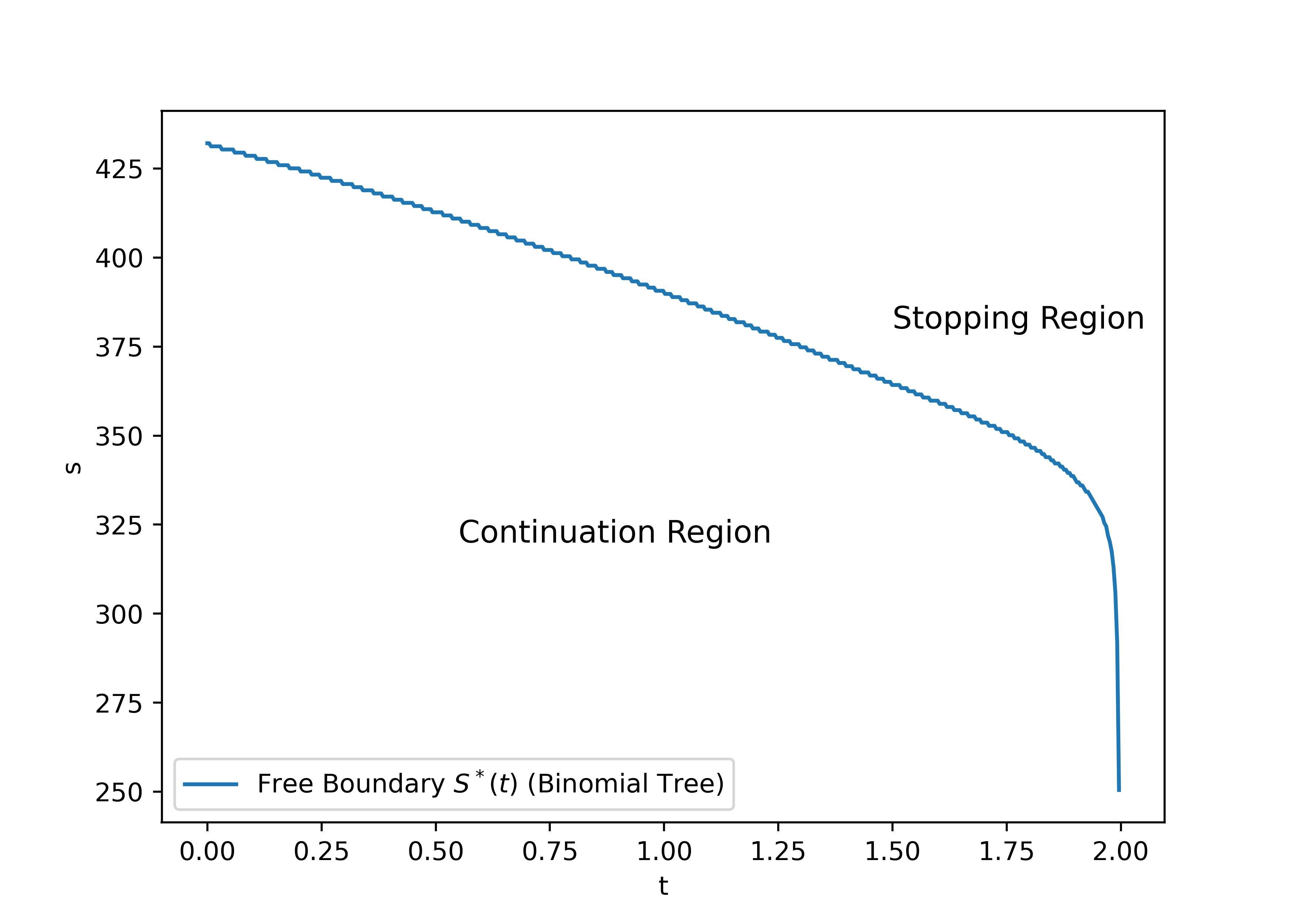}}
\caption{Free boundary for American options obtained by binomial tree method. $(a),$ Free boundary for American put option with $K = 100, r = 0.02, \sigma = 0.25, T = 1, q = 0$. $(b),$ Free boundary for American call option with $K = 200, r = 0.05, \sigma = 0.25, T = 2, q = 0.03$.}
    \label{fig6}
\end{figure}

For American call options, a free boundary is similarly defined. Unlike put options, the region above the boundary represents the stopping region, while the region below corresponds to the continuation region.
\begin{align}
    \mathcal{S} = \{s \geq S^*(t), 0 \leq t \leq T\} &= \{(s,t):V(s,t) = (s-K)^+\},  \notag\\ \mathcal{C} = \{0 \leq s < S^*(t), 0 \leq t \leq T\} &= \{(s,t):V(s,t) > (s-K)^+\},   \notag\\ S^*(t) &= \inf \{s\in \mathbb{R}^+: V(s, t) = (s-K)^+ \}.  \label{18''}
\end{align}
We plot the free boundary for both American put and call options in Figure~\ref{fig6}, using the solution obtained from the binomial tree method with 
$N = 4000$ as the reference solution. The free boundary $S^*(t)$ is then calculated by Eq.~(\ref{15}) and (\ref{18''}).
The uncertainty of $S^*(t)$ turns it into a free boundary problem, making the pricing of American options more complex than European options. Eq.~(\ref{17}) - (\ref{18'}) has no explicit analytical solutions and must therefore be solved by numerical methods.

\subsubsection{High-Dimensional BSM Equations for Multi-Asset Options}
\label{sec2.1.4}

In addition to single-asset options, the BSM model can be extended to high-dimensional multi-asset models. This section focuses on the BSM framework for options on $n$ underlying assets. The price dynamics of these assets are described by the following $n$-dimensional geometric Brownian motions in the risk-neutral form \cite{kovalov2007pricing},  \begin{equation}
    dS_i(t) = (r - q_i)S_i(t)dt + \sigma_iS_i(t)dW_i, \label{55}
\end{equation}
where $r$ is the risk-free rate, $\sigma_i$ is the volatility of the $i$-th asset, and $q_i$ is the dividend yield. $T$ is the expiration date. $\{W_i\}$ are $n$ standard Brownian motions with instantaneous coefficients of correlation denoted by $\rho_{ij}$, \begin{equation}
    dW_idW_j = \rho_{ij}dt.  \label{56}
\end{equation}

We consider an American option with a payoff function $\Phi = \Phi(S_1(t),\cdots, S_n(t))$.
Then the value of this American option of $n$ underlying assets $V = V(s_1, \cdots, s_n, t)$ satisfies the following BSM model \cite{zhang2015efficient, bustamante2016multi},
\begin{numcases}{}
\quad \mathcal{F}\big(V(s_1, \cdots, s_n, t)\big) \leq 0,\quad  \forall s_i \geq 0 ,    t \in [0, T), \notag \\ \quad 
 \mathcal{TV}\big(V(s_1, \cdots, s_n, t)\big) \geq 0, \quad  \forall s_i \geq 0 ,  t \in [0, T),  \notag \\ \quad \mathcal{F}\big(V(s_1, \cdots, s_n, t)\big) \cdot \mathcal{TV}\big(V(s_1, \cdots, s_n, t)\big) = 0, \quad  \forall s_i \geq 0 ,   t \in [0, T),  \notag \\ \quad V(s_1, \cdots, s_n, T) = \Phi(s_1, \cdots, s_n),  \quad \forall s_i \geq 0, \notag 
\end{numcases}
where in the multi-asset cases, the operators are defined as follows, \begin{align}
    \mathcal{F}(V(s_1, \cdots, s_n, t)) &= \frac{\partial V}{\partial t} + \frac{1}{2} \sum \limits_{i,j=1}^n\sigma_i \sigma_j \rho_{ij}s_is_j \frac{\partial^2 V}{\partial s_i \partial s_j} + \sum \limits_{i=1}^n (r-q_i)s_i\frac{\partial V}{\partial s_i}-rV.  \label{26}  \\ \mathcal{TV}\big(V(s_1, \cdots, s_n, t)\big) &= V(s_1, \cdots, s_n, t) - \Phi(s_1, \cdots, s_n). \notag 
\end{align}

In practice, there are various types of American multi-asset options with different payoff functions $\Phi(s_1, \cdots, s_n)$. Common types of payoffs include options based on the maximum or minimum of several asset prices, spread options that consider the difference between two prices, and portfolio options that average prices. Contingent claims with these features are prevalent in financial exchanges, over-the-counter transactions, cash flows resulting from corporate investment decisions, and executive compensation plans \cite{detemple2005american}. Table~\ref{table0} provides examples of some commonly used American multi-asset options.

\begin{table}[!ht]
\centering
\caption{Different types of multi-asset options and their payoff functions}
\label{table0}
\renewcommand{\arraystretch}{1.5} 
\begin{tabular}{cc|cc}
\hline
Type of option & $\Phi(s_1, \cdots, s_n)$  & Type of option & $\Phi(s_1, \cdots, s_n)$\\ \hline 

Call-on-max \cite{detemple2003valuation} & $\big [ \max \{s_i\}_{i=1}^n-K\big ]^+$ &
Call on geometric mean \cite{eytan1986pricing}& $\big [(\prod\limits_{i=1}^{n} s_i)^{\frac{1}{n}}-K\big ]^+$ \\

Call-on-min \cite{detemple2003valuation} & $\big [\min \{s_i\}_{i=1}^n-K\big ]^+$ &
Call on arithmetic mean \cite{sirignano2018dgm} & $\big [(\frac{1}{n}\sum\limits_{i=1}^ns_i)-K\big ]^+$ \\

Put-on-max \cite{chan2006pricing} & $\big [K-\max \{s_i\}_{i=1}^n\big ]^+$ &
Put on geometric mean \cite{kovalov2007pricing}& $\big [K-(\prod\limits_{i=1}^{n} s_i)^{\frac{1}{n}}\big ]^+$ \\
	
Put-on-min \cite{rogers2002monte} & $\big [K-\min \{s_i\}_{i=1}^n\big ]^+$ & 
Put on arithmetic mean \cite{jin2016efficient} & $\big [K-(\frac{1}{n}\sum \limits_{i=1}^ns_i)\big ]^+$ \\\hline 
\end{tabular}
\end{table}

\subsection{Physics-Informed Neural Networks}
This section introduces the concept of Physics-Informed Neural Networks (PINN) \cite{RAISSI2019686} and discusses their modification for PDEs subject to inequality constraints, as encountered in the pricing of American options.

\subsubsection{General Framework}

In the context of PINN, partial differential equations in the following form are usually considered 
\begin{equation*}
    \mathcal{G}(u) = u_t + \mathcal{N}(u) = 0, \quad \forall x \in \mathcal{D},  t \in [0, T),   
\end{equation*}
subject to the initial and boundary conditions
\begin{equation*} 
\left\{  
\begin{aligned} &\quad
   u(x, 0) = g(x), \quad \forall x \in \mathcal{D}, \\ &\quad \mathcal{B}(u) = 0, \quad \forall x \in \partial \mathcal{D},  t \in [0, T], 
   \end{aligned}  
\right.    
\end{equation*}
where $\mathcal{N}[\cdot]$ is a differential operator, $\mathcal{B}[\cdot]$ is a boundary operator representing Dirichlet, Neumann, Robin, or periodic boundary conditions, and $\mathcal{D}$ is a subset of $\mathbb{R}^d$. The purpose is to approximate the unknown solution $u(x, t)$ by a deep neural network $u_{\theta}(x, t)$, referred to as PINN, where $\theta$ is the parameter of the network. The model can be trained by minimizing the following composite loss function, 
\begin{equation*}
    \mathcal{L}(\theta) = \lambda_f \mathcal{L}_f(\theta) + \lambda_{ic}\mathcal{L}_{ic}(\theta) + \lambda_{bc}\mathcal{L}_{bc}(\theta). 
\end{equation*}
Each term is specifically defined as follows, \begin{align}  \mathcal{L}_f(\theta) &= \frac{1}{N_f}\sum \limits_{i=1}^{N_f} |\mathcal{G}(u_{\theta}(x_f^i, t_f^i))|^2,   \notag \\
    \mathcal{L}_{ic}(\theta) &= \frac{1}{N_{ic}}\sum \limits_{i=1}^{N_{ic}} |u_{\theta}(x_{ic}^i, 0) - g(x_{ic}^i)|^2,   \label{13'}\\
    \mathcal{L}_{bc}(\theta) &= \frac{1}{N_{bc}}\sum \limits_{i=1}^{N_{bc}} |\mathcal{B}(u_{\theta}(x_{bc}^i, t_{bc}^i))|^2.  \notag 
\end{align}
Here $\{x_f^i, t_f^i\}_{i = 1}^{N_f}$ denotes the collocations points for $\mathcal{G}(u)$. $\{x_{ic}^i, 0\}_{i = 1}^{N_{ic}}$ and $\{x_{bc}^i, t_{bc}^i\}_{i = 1}^{N_{bc}}$ are collocations points for initial and boundary conditions. $\{\lambda_f, \lambda_{ic}, \lambda_{bc}\}$ are hyperparameters that assign  weights to different loss terms.

\subsubsection{Loss Design for PDEs with Inequalities}

BSM equations for European options only contain equality constraints, such as Eq.~(\ref{9}). Therefore, the loss function adopts a structure similar to that of PINN,\begin{equation*}
    \mathcal{L}(\theta) = \lambda_f \mathcal{L}_f(\theta) + \lambda_{tc}\mathcal{L}_{tc}(\theta).   
\end{equation*}
Here $\mathcal{L}_{tc}$ represents the terminal condition term, which is similar to Eq.~(\ref{13'}), \begin{equation*}
    \mathcal{L}_{tc}(\theta) =\frac{1}{N_{tc}}\sum \limits_{i=1}^{N_{tc}} \Big|(u_{\theta}(x_{tc}^i, T) - \Phi(x_{tc}^i))\Big|^2.   
\end{equation*}

To address the challenges posed by inequalities in the system of Eq.~(\ref{17})-(\ref{18'}) for American options, we modify the loss function as follows, 
\begin{equation*}
    \mathcal{L}(\theta) = \lambda_{bs} \mathcal{L}_{bs}(\theta) + \lambda_{tv}\mathcal{L}_{tv}(\theta) + \lambda_{eq}\mathcal{L}_{eq}(\theta) + \lambda_{tc}\mathcal{L}_{tc}(\theta),    
\end{equation*}
where $\{\lambda_{bs}, \lambda_{tv}, \lambda_{eq}, \lambda_{tc}\}$ are hyperparameters that assign weights to different loss terms. The first term $\mathcal{L}_{bs}$ is defined as \begin{equation*}
    \mathcal{L}_{bs}(\theta) = \frac{1}{N_{bs}}\sum \limits_{i=1}^{N_{bs}}  \Big|\max\Big(\mathcal{F}\big(u_{\theta}(s_{bs}^i, t_{bs}^i)\big), 0\Big)\Big|^2.  
\end{equation*}
where $\mathcal{F}(V(s, t))$ is defined in Eq.~(\ref{9'}) for single-asset options and Eq.~(\ref{26}) for multi-asset options. $s_{bs}^i$ is the vector of underlying asset prices at the $i$-th collocation point. Since the linear complementarity conditions require $\mathcal{F}(V) \leq 0$, the term $\mathcal{L}_{bs}$ serves as a penalty to enforce this constraint. Specifically, $\mathcal{L}_{bs}$ imposes a penalty whenever the predicted value $u_{\theta}$ results in $\mathcal{F}(u_{\theta}) > 0$. When $\mathcal{F}(u_{\theta}) \leq 0$, this term degenerate to $0$.

The time value of American options should be nonnegative. Consequently, the loss term associated with this constraint is defined as follows, 
\begin{equation*}
    \mathcal{L}_{tv}(\theta) = \frac{1}{N_{tv}}\sum \limits_{i=1}^{N_{tv}} \Big|-\min \Big(\mathcal{TV}\big(u_{\theta}(s_{tv}^i, t_{tv}^i)\big), 0 \Big)\Big|^2.  
\end{equation*}
Similar to $\mathcal{L}_{bs}$, this term acts as a penalty term to enforce the non-negativity of the time value, ensuring it remains greater than or equal to zero. This term degenerates to $0$ when this inequality is satisfied.
To account for the equality in the linear complementarity conditions, we introduce a third term in the loss function defined, 
\begin{equation*}
   \mathcal{L}_{eq}(\theta) =  \frac{1}{N_{eq}}\sum \limits_{i=1}^{N_{eq}} \Big|\mathcal{F}\big(u_{\theta}(s_{eq}^i, t_{eq}^i)\big) \cdot \mathcal{TV}\big(u_{\theta}(s_{eq}^i, t_{eq}^i)\big)\Big|^2.
\end{equation*}
This loss term is designed to ensure that at least one of the inequalities is satisfied as an equality.
By weighting and summing these four terms, the loss function for solving the system of inequality equations is obtained.

\section{Methodology and Formulation}

\label{sec3}

In this section, we present the methodological framework of the proposed Exact Terminal Condition Neural Network (ETCNN). Section~\ref{sec3.1}  introduces the core idea of the ETCNN approach, followed by a description of the network architecture in Section~\ref{sec3.2} and the input normalization strategy in Section~\ref{sec3.3}. Section~\ref{sec3.4} applies the proposed framework to a European option, serving as a toy model to illustrate the fundamental mechanism of the method. Finally, Section~\ref{sec3.5} extends the formulation to American options, laying the foundation for the numerical experiments presented in Section 4.

\subsection{Imposing Exact Terminal Conditions}
\label{sec3.1}

As we will show in Figure \ref{fig9}, the standard PINN method suffers from low solution accuracy, partly due to the non-differentiability of terminal conditions at specific points and the near-singular behavior close to the terminal. To address these limitations, we propose the ETCNN in this section.

The core idea of imposing the exact terminal conditions method is to choose a trial function that automatically satisfies the terminal condition to approximate the solution. The idea of embedding exact boundary conditions into neural network trial functions has been considered in prior studies \cite{berg2018unified,lagaris1998artificial}, and we build upon this line of work by introducing a tailored design for the BSM equations. Consider a system of differential equations formulated as follows:
\begin{equation} 
\left\{  
\begin{aligned} 
\quad &
   \mathcal{G}(V(s,t)) = 0, \quad \forall  s \in \mathcal{D},   t \in [0, T), \\ 
\quad &  
   V(s, T) = \Phi(s), \quad \forall s \in \mathcal{D}. \label{40}
   \end{aligned}
\right.   
\end{equation}
where $t$ is the current time, $s$ is a scaler or a vector defined on a region $\mathcal{D}$. The equality sign in the first line is not necessarily required and can be an inequality, similar to the BSM equation for American options. This equation is subject to some terminal condition function $\Phi(s)$.

The trial solution is chosen as 
\begin{equation*}
    \Tilde{u}_{NN}(s, t) = g_1(s,t) u_{NN}(s,t) + g_2(s, t). 
\end{equation*}
where $g_1, g_2$ are constructed to satisfy specific constraints,
\begin{equation*}
    g_1(s, T) = 0, \quad g_2(s, T) = \Phi(s), \quad \forall s \in \mathcal{D}. 
\end{equation*}
By this construction, the overall trial solution $\Tilde{u}_{NN}$ inherently satisfies the terminal conditions.  

The main advantage of this approach is that the resulting solution exactly satisfies the boundary conditions, thereby eliminating the errors in boundary condition enforcement that are typical of conventional PINNs and consequently improving solution accuracy. Another advantage is that our designed $g_2$ is constructed to preserve key properties of the true solution.
In addition to satisfying the terminal conditions, it is desirable for $g_2$ to ensure that the trial solution captures key properties of the true solution, such as smoothness and singular behaviors. True solutions are often non-differentiable at some points on the terminal but are smooth anywhere else. They also exhibit near-singularity near the terminal. Therefore, it is generally inappropriate to simply set $g_2(s,t) = \Phi(s)$, as this function neither preserves the differentiability nor captures the singularities present in the true solution. As demonstrated in 
Section $\ref{Sec4.1}$, poor choice of $g_2$ may even reduce the accuracy of the network. In the subsequent sections, we will discuss how to appropriately select $g_2$ for the BSM equations for both European and American options.

\subsection{Neural Network Structure}
\label{sec3.2}
In the context of applying deep learning to solve partial differential equation 
problems, residual networks (ResNet) are commonly used for function approximation. Deep residual learning, introduced by He et al. \cite{he2016deep}, is a neural network architecture based on the concept of residual blocks. Each residual block consists of several fully connected layers, where the final output is computed by adding the input of the block to the output of its last layer.
In this section, we provide an overview of the ResNet structure and explain how it is adapted to satisfy accurate terminal conditions.

Consider a ResNet containing $M$ residual blocks, each consisting of $L$ layers with $n$ neurons per layer.
When solving Eq.~(\ref{40}), suppose the input vector $x = (s, t)$ has a dimension $d_{in}$. The network begins with a fully connected layer, transforming the input into an $n$-dimensional vector, \begin{equation*}
    g^{(1,0)}(x) = \sigma(W^{in}\cdot x + b^{in}),
\end{equation*}
where $W^{in} \in \mathbb{R}^{n\times d_{in}}$ is a weight matrix, $b^{in} \in \mathbb{R}^n$ is a bias vector and $\sigma$ is the activation function. We use the $\tanh$ function in this work. Let $g^{(m, 0)}(x)$ be the input of the $m$-th block, the structure of the $m$-th block is defined as follows, \begin{equation*}
    f_{\theta}^{(m,l)}(x) = W^{(m,l)}\cdot g^{(m,l-1)}(x) + b^{(m,l)}, \quad g^{(m,l)}(x) = \sigma(f_{\theta}^{(m,l)}(x)), \quad 1 \leq l \leq L, 1 \leq m \leq M.  
\end{equation*}
Here, $W^{(m,l)} \in \mathbb{R}^{n \times n}$ is the weight matrix in the $l$-th layer. $b^{(m,l)} \in \mathbb{R}^n$ is the bias vector. The final output of the $m$-th block, which is the input of the next part, is 
\begin{equation*}
    g^{(m+1, 0)}(x) = f_{\theta}^{(m,L)}(x) + g^{(m,0)}(x), \quad 1 \leq m \leq M   
\end{equation*}

\begin{figure}[!h]
    \centering  \includegraphics[width=0.75\textwidth]{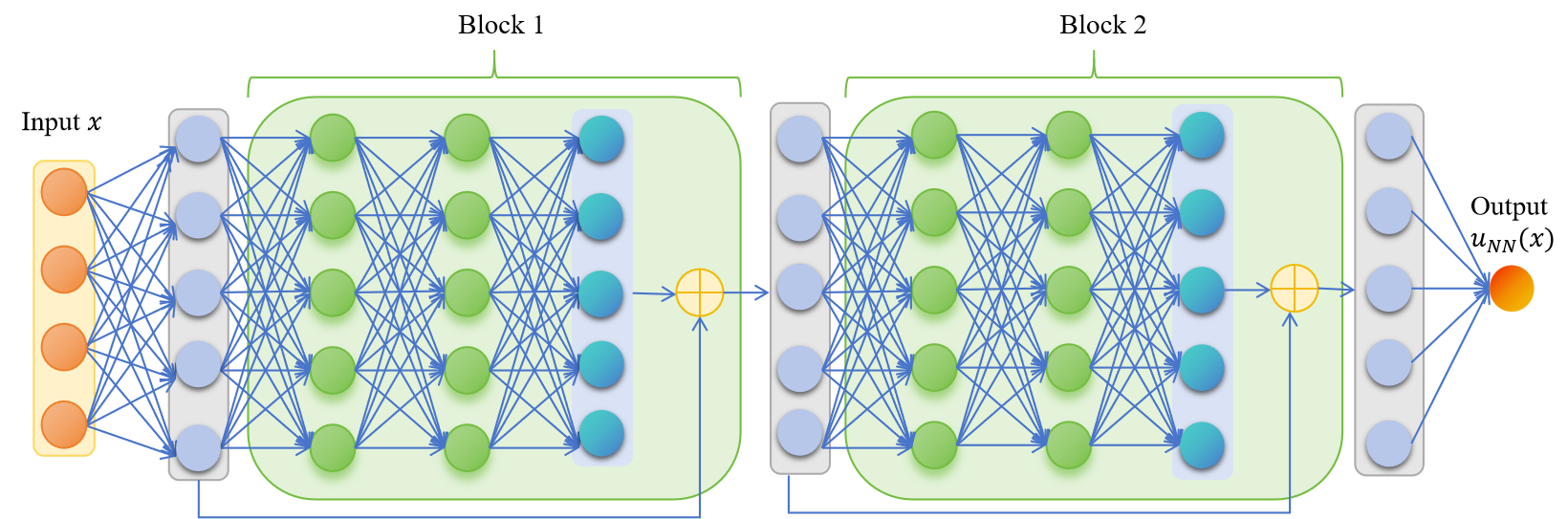}\\
    \caption{Structure of a ResNet with $M = 2, L = 3, n = 5, d_{in} = 4, d_{out} = 1$.}
        \label{fig3}
\end{figure}

Finally, the output of the network is \begin{equation}
    f_{\theta}(x) = W^{out}\cdot g^{(M+1, 0)}(x) + b^{out}. \label{16'}
\end{equation}
Here, $W^{out} \in \mathbb{R}^{d_{out} \times n}, b^{out} \in \mathbb{R}^{d_{out}}$, where $d_{out}$ is the dimension of solution. Figure~\ref{fig3} illustrates the structure of a ResNet with parameters $M = 2, L = 3, n = 5, d_{in} = 4$ and $d_{out} = 1$.

\begin{figure}[!h]
    \centering  \includegraphics[width=0.5\textwidth]{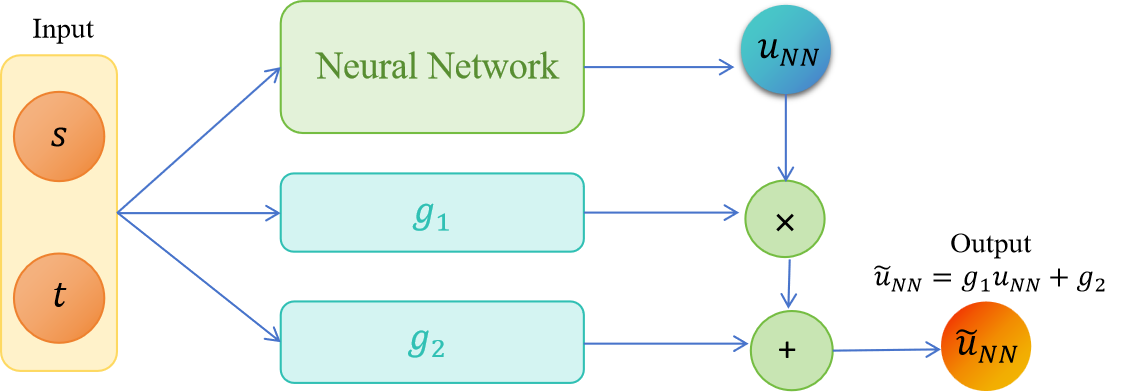}\\
    \caption{Structure of our exact terminal method.}
        \label{fig4}
\end{figure}

In the exact terminal method, the output layer of the network is modified to ensure that the terminal conditions are satisfied, in a manner similar in \cite{guo2024deep}.
The network architecture in front of the output layer remains unchanged, only the last layer is modified. 
For ResNet, Eq.~(\ref{16'}) is modified as 
\begin{equation*}
    f_{\theta}(x) = g_1(x) \cdot (W^{out}\cdot g^{(M+1, 0)}(x) + b^{out}) + g_2(x). 
\end{equation*}
Here, $g_1$ and $g_2$ are functions that need to be assigned before training which satisfy the following conditions, \begin{equation*}
     g_1(x) = 0, \quad g_2(x) = \Phi(s), \quad \forall x = (s, T) \in \mathcal{D} \times \{T\}. 
\end{equation*}
Figure~\ref{fig4} shows the structure of our ETCNN.

\subsection{Input Normalization}
\label{sec3.3}
We further investigate normalizing the input variables by transforming the asset price $s$ into its moneyness, defined as the ratio $s/K$. This normalization ensures that the input variables are of similar orders of magnitude, as the time variable $t$ is typically of order $\mathcal{O}(1)$ since the term of an option is usually a few months or years. 
A more significant reason to do such normalization lies in the homogeneity property of the option pricing function $V$ with respect to $s$. When the option pricing formula $V$ is expressed as a function of asset price, strike price and time, it exhibits a homogeneous property \cite{merton1973theory},
\begin{equation*}
    \alpha \cdot V(s, K, t) = V(\alpha s, \alpha K,  t)  
\end{equation*}
This property can be easily derived from the characteristics of geometric Brownian motion. For the options under our consideration, a fixed strike price $K$ is predetermined. Consequently, the solution depends on the ratio $s/K$. 

\begin{figure}[!h]
    \centering  \includegraphics[width=0.6\textwidth]{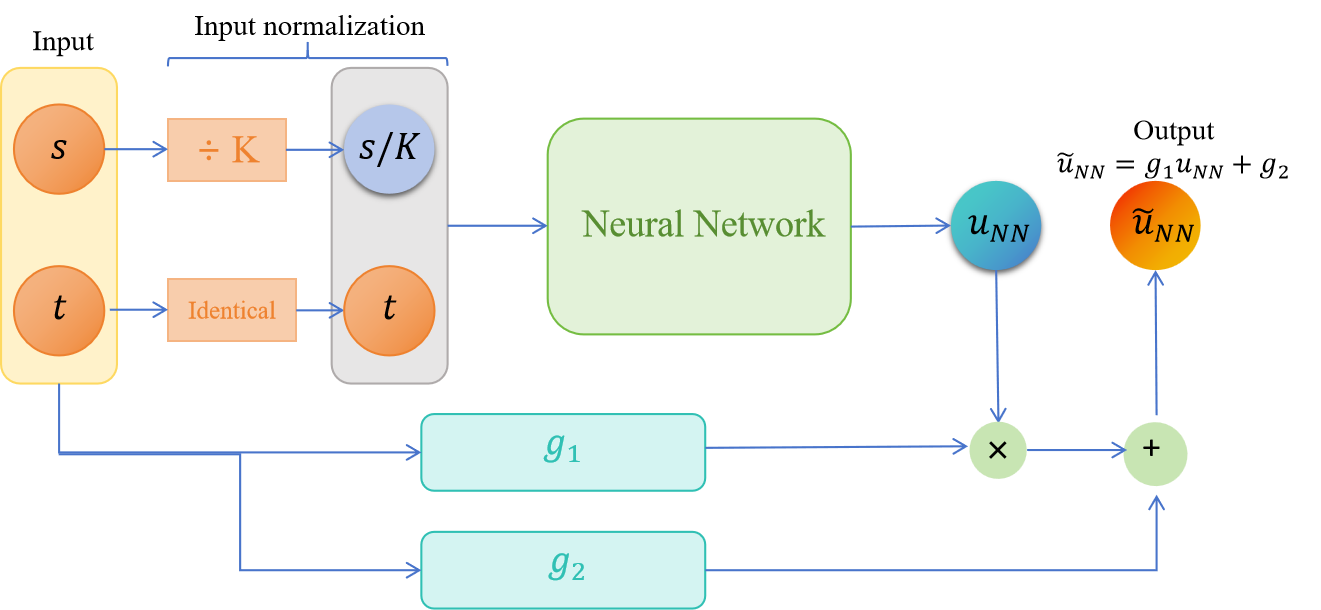}\\
    \caption{Structure of our exact terminal method with input normalization.}
        \label{fig5}
\end{figure}

The input normalization structure introduced in this section is designed to maintain this property for the neural network solution. When the input vector $(s, t)$ enters the network, the asset price vector $s$ is first normalized by dividing each of its dimensions by $K$, while the temporal component $t$ remains unchanged. The normalized input $(s/K, t)$ is then passed into the network. In this way, the output of the network will be a function of $s/K$. Figure~\ref{fig5} shows the overall structure of our ECTNN with an input normalization layer.

\subsection{Illustrative Example: European Option} 
\label{sec3.4}
This example serves as a toy model for solving a basic BSM equation for pricing single-asset European options, which are the simplest type of option. We use this experiment to demonstrate the effectiveness of our ETCNN, to assess which network structures are most suitable, and to explore the criteria necessary for selecting exact terminal functions. These criteria will later be applied to provide exact terminal functions for American options, which are the primary focus of this article. Meanwhile, the analytical solution for European options will serve as the foundation for determining exact terminal functions for American options in the next subsections. 

Single-asset European call options satisfy the following BSM equation,
\begin{equation*} 
\left\{  
\begin{aligned} \quad &
   \mathcal{F}(V) = \frac{\partial V}{\partial t} + \frac{1}{2}\sigma^2 s^2 \frac{\partial^2 V}{\partial s^2} + rs\frac{\partial V}{\partial s}-rV = 0,  \quad \forall s \geq 0,   t \in [0, T),  \\ \quad &
   V(s, T) =  (s-K)^+, \quad \forall s \geq 0. 
   \end{aligned}
\right.     
\end{equation*}
This equation has an explicit solution,
\begin{equation}
    V(s,t) = s \cdot N\big(d_1(s, \tau, K)\big) - Ke^{-r\tau}\cdot N\big(d_2(s, \tau, K)\big), \label{54}
\end{equation}
where $\tau = T-t$ is the time to maturity. $N(\cdot)$ is the cumulative distribution function of the standard normal distribution. $d_1, d_2$ are defined as follows, 
\begin{equation}
 d_1(s, \tau, K) = \frac{1}{\sigma \sqrt{\tau}}\big(\ln{\frac{s}{K}}+(r+\frac{\sigma^2}{2})\tau\big), \quad
   d_2(s, \tau, K) =  d_1(s, \tau, K) - \sigma \sqrt{\tau}=\frac{1}{\sigma \sqrt{\tau}}\big(\ln{\frac{s}{K}}+(r-\frac{\sigma^2}{2})\tau\big).   \label{13}
\end{equation}
When $\tau = 0$, we define \begin{equation*}
    N\big(d_1(s, 0, K)\big) = N\big(\lim \limits_{\tau \to 0}d_1(s, \tau, K)\big) = \begin{cases}
1, &\text{if}\ s > K,\\
\frac{1}{2}, &\text{if}\ s = K, \\ 
0, &\text{if}\ s < K.
\end{cases}  
\end{equation*}

We take $K = 100, r = 0.05, \sigma = 0.15, T = 1$ in this experiment. The strike price of options traded on exchanges usually does not deviate too much from the underlying assets' prices. In this case, the network is trained on $[50, 150] \times [0, T]$ and evaluated on $[80, 120] \times [0, T]$, which covers the vast majority of situations in actual transactions. The training domain is set slightly wider than the testing domain to ensure reliable accuracy within the region of practical interest.

The terminal condition in this case is \begin{equation*}
    \Phi(s) = (s-K)^+ = \frac{1}{2}(s - K + \sqrt{s^2 + K^2-2sK}).  
\end{equation*}
To construct an appropriate $g_2$, we slightly modify $\Phi(s)$ by introducing the time variable $t$ and
obtain a function on $(s, t)$ that both satisfies the terminal conditions at $t = T$ and is a differentiable function when $t < T$. A common technique in the analysis of European options at time $t$ is to discount strike price $K$ at a discount rate $r$ to time $t$. Inspired by this, we naturally extend the idea of discounting $K$ and define
\begin{equation*}
    g_1(s, t) = T-t, \quad g_2(s, t) = \frac{1}{2}(s - K + \sqrt{s^2+K^2-2sKe^{-r(T-t)}}).  
\end{equation*}
With this construction, the solution $\Tilde{u}_{NN} = g_1\cdot u_{NN} + g_2$ accurately satisfies the terminal condition $\Tilde{u}_{NN}(s, T) = (s-K)^+$. Furthermore, $\Tilde{u}_{NN}$ retains non-differentiability at the point $(K, T)$ while remaining smooth elsewhere. This design ensures that $\Tilde{u}_{NN}$ matches the differentiability properties of the true solution. 

In this case, and throughout all experiments in Section~\ref{sec4}, the Adam \cite{kingma2017adammethodstochasticoptimization} optimization algorithm is employed to obtain the network parameters $\theta$. The hyperparameter $\beta$ in Adam denotes the pair of exponential decay rates $(\beta_1, \beta_2)$ for the first- and second-moment estimates. In our implementation, we adopt the standard setting $(\beta_1, \beta_2) = (0.9, 0.999)$.  The learning rate $lr$ starts at an initial value $lr_{start}$ and decays exponentially with a decay factor $\gamma$. To be specific, each of our experiments involves 200,000 training iterations. During the first 40,000 iterations, the learning rate decays by a factor of $\gamma$ every 2,000 iterations. In the remaining 160,000 iterations, the decay occurs every 5,000 iterations. This two-stage schedule was empirically designed: the initial frequent decays help the network quickly reach a reasonable solution, while the subsequent slower decay pace allows for more precise convergence in the later stages of optimization.
Both $lr_{start}$ and $\gamma$ are hyperparameters tuned for each experiment. Details of the hyperparameters $\{\lambda\}$, $lr_{start}$ and $\gamma$ 
will be provided in the context of individual experiments.

For the toy model under consideration, the parameters are set as $\lambda_f = 20, \lambda_{tc} = 1$. For loss computation, $N_{tc} = 1024$ points are sampled to calculate $\mathcal{L}_{tc}$, and $N_{f} = 4N_{tc}$ to calculate $\mathcal{L}_{f}$. Notably, ETCNN eliminates the need for tuning $\lambda_{tc}$. For consistency, ETCNN uses the same value of $\lambda_f$ and $N_f$.  The learning rate is initialized at $lr_{start} = 0.01$, with a decay factor of $\gamma = 0.85$. 
To evaluate the effectiveness of ETCNN and the impact of input normalization, we employ ResNet architectures with four and five blocks, where each block consists of two layers, and each layer contains 50 neurons. Both configurations are tested with and without the input normalization layer. The performance is assessed by the $L^2$ relative errors between the computed solutions and the exact solution. Each experiment is repeated three times, and the average results are reported. The outcomes are summarized in Table \ref{table1}.

The results demonstrate that applying exact terminal conditions significantly enhances accuracy, improving it by approximately one order of magnitude. Furthermore, combining normalization with exact terminal conditions yields the best results. The performance of the 5-block network is comparable to that of the 4-block network. However, the 4-block network requires fewer parameters. This suggests that a 4-block network may represent an optimal choice for the number of blocks. Based on these findings, we adopt a 4-block ResNet for subsequent numerical experiments.

\begin{table}[!h]
    \centering
    \caption{Relative $L^2$ error on PINN and ETCNN. Norm refers to the input normalization. Bold font represents the best result in each row.}
 \label{table1}
\renewcommand{\arraystretch}{1.5} 
    \begin{tabular}{c|cccc}
    \hline
        ResNet & PINN & PINN+Norm & ETCNN & ETCNN+Norm \\ \hline
        4-block & $2.13\times 10^{-3}$ & $1.65\times 10^{-3}$ & $4.63\times 10^{-4}$ & $\textbf{2.94}\times \textbf{10}^{\textbf{-4}}$ \\ 
        5-block & $2.33\times 10^{-3}$ & $1.68\times 10^{-3}$ & $4.37\times 10^{-4}$ & $\textbf{3.83}\times \textbf{10}^{\textbf{-4}}$ \\ 
 \hline
    \end{tabular}
\end{table}

\subsection{Methodological Formulation for American Options} 
\label{sec3.5}

The American BSM equation is more complex to solve due to inequality constraints. In order to improve the accuracy of the solution, it is necessary to carefully select $g_2$. We propose the exact terminal conditions methods based on our understanding of the financial properties of options, and by borrowing knowledge from European options. It is natural to view the price of American options as the sum of the price of an equivalent European option and an early exercise premium.
Several papers have provided a theoretical basis for this \cite{kim1990analytic, carr1992alternative, kitapbayev2021closed}. 
\begin{equation*}
    V^a(s, t) = V^e(s, t) + p(s, t),  
\end{equation*}
where $s \in \mathbb{R}^n$ is the price of $n$ underlying assets. $V^a(s,t)$ represents the price of the American option. $V^e(s,t)$ is the value of its corresponding European option, which shares the same parameters as the American option but can only be exercised at the expiration date. $p(s, t)$ is the early exercise premium. 

Klimsiak and Rozkosz \cite{klimsiak2016early} provided an implicit integral representation for
the premium $p(s, t)$ under specific assumptions.
If the terminal payoff $\Phi$ is a nonnegative continuous function and is smooth on $\{V^a(s,t) = \Phi(s)\} \cap \mathbb{R}^n \times [0,t]$ for every $t \in (0,T)$, or $\Phi$ is a nonnegative convex function, then \begin{equation}
    p(s, t) = \mathbf{E}_{t, s}\left[\int_t^Te^{-r(u-t)}\textit{I}\{V^a(s(u),u)=\Phi(s(u))\}H(s(u),u)du\right], \label{58}
\end{equation}
where $\mathbf{E}_{t, s}$ is the expectation under risk-neutral measure and $\textit{I}(\cdot)$ is the indicator function. $H(s, t) = -\mathcal{F}(\Phi(s(u),u))$ where $\mathcal{F}$ is the PDE operator defined in Eq.~(\ref{26}). The region $\{V^a(S(u),u)=\Phi(S(u))\}$ corresponds to the stopping region and is unknown since the free boundary $S^*(t)$ is unknown.

Note that $V^e(s,t)$ inherently satisfies the terminal condition, as the corresponding European option shares the same terminal condition. This observation naturally leads to the idea of extracting information from $V^e(s,t)$ to construct $g_2(s, t)$.
Furthermore, $p(s,t)$ involves an integral over time. If $p(s,t)$ does not include singular integrals or contains nearly singular terms with minimal impact after integration, then the singularities in $V^a(s,t)$ primarily originate from $V^e(s,t)$. Consequently, we design a function $g_2(s,t)$ that approximates $V^e(s,t)$, such that $g_2$ satisfies exact boundary conditions while also incorporating the singularity in $V^e$. This way, the remaining residue $V^a-g_2$ becomes smooth, making it easier for the neural network to learn.
The subsequent section provides detailed examples and the explicit construction of $g_2$.

\section{Numerical Results for American Options}
\label{sec4}

To illustrate the broad applicability of our method, we consider four distinct examples in this section, each with different settings and parameters. Section~\ref{Sec4.1} presents a single-asset American put option, while Section~\ref{Sec4.2} studies single-asset American call options with dividends, including variations in volatility and dividend rates. We provide comparisons with traditional numerical methods for these two examples. Section~\ref{Sec4.3} examines multi-asset options with a geometric-average payoff, considering different numbers of underlying assets, and Section~\ref{Sec4.4} focuses on a two-asset call-on-max option with varying volatilities and dividend rates. Comparisons with traditional methods are not included in these two high-dimensional examples, as they become computationally infeasible. The parameters such as $K, r$, and $\rho$ are chosen differently in these examples. By presenting this variety of examples with differing parameters and characteristics, we aim to demonstrate that the proposed method is robust and effective across a wide range of option types, dimensionalities, and market conditions.  

For all experiments in this section, we employ a ResNet architecture consisting of $M = 4$ blocks, where each block contains $L = 2$ layers with $50$ nodes per layer. The hyperparameters are set as follows, $\lambda_{bs} = \lambda_{tv} = \lambda_{eq} = \lambda_{tc} = 1, lr_{start} = 0.01, \gamma = 0.9$. The ETCNN takes the same $\lambda_{bs}, \lambda_{tv}$ and $\lambda_{eq}$. Additionally, the number of sampling points doubles every 80,000 iterations. Network architecture and hyperparameters were chosen based on standard practices in the literature and our own preliminary tests. Each experiment in this study is repeated three times, and the average results are reported to mitigate the stochastic effects of neural network training and ensure result reliability. All experiments are conducted on an NVIDIA A800 GPU.

\subsection{Single-Asset American Put Options}

\label{Sec4.1}
This example considers a single-asset non-dividend-paying American put option. The value of the option  $V(s, t)$ satisfies the following complementarity conditions, 
\begin{subequations}
\begin{numcases}{}
\quad  V(s, t) \geq (K-s)^+,\quad  \forall s \geq 0, t \in [0, T),  \\ \quad  
\frac{\partial V}{\partial t} + \frac{1}{2}\sigma^2 s^2 \frac{\partial^2 V}{\partial s^2} + rs\frac{\partial V}{\partial s}-rV \leq 0,\quad  \forall  s \geq 0,t \in [0, T),\\ \quad \big(\frac{\partial V}{\partial t} + \frac{1}{2}\sigma^2 s^2 \frac{\partial^2 V}{\partial s^2} + rs\frac{\partial V}{\partial s}-rV  \big)\cdot \big(V-(K-s)^+ \big)=0,\quad  \forall  s \geq 0, t \in [0, T), \\ \quad  V(s, T) = (K-s)^+.  \label{66}
\end{numcases}
\end{subequations}
The term $(K-s)^+$ denotes the terminal payoff. We design some terminal functions $g_2$ with specific properties to promote solution accuracy and stability.

\subsubsection{Exact Terminal Function Design}

As previously stated, the value of an American put option can be written as
\begin{equation*}
    V^a(s, t) = V^e(s, t) + p(s, t),  
\end{equation*}
where $V^e(s, t)$ is the value of the corresponding European put. For simplicity of symbols, denote $\Tilde{d_1}$ and $\Tilde{d_2}$ as\begin{equation*}
    \Tilde{d_1}(s, \tau, K)  = -\frac{1}{\sigma \sqrt{\tau}}\big(\ln \frac{s}{K} + (r + \frac{\sigma^2}{2})\tau\big),  \quad  \Tilde{d_2}(s, \tau, K)  = -\frac{1}{\sigma \sqrt{\tau}}\big(\ln \frac{s}{K} + (r - \frac{\sigma^2}{2})\tau\big).  
\end{equation*}
The value of the European option is given by  
\begin{equation}
V^e(s, t) = Ke^{-r\tau}\cdot N\big(\Tilde{d_2}(s, \tau, K)\big) - s\cdot N\big(\Tilde{d_1}(s, \tau, K)\big). \label{16a}
\end{equation}

The term $p(s, t)$ is the early exercise premium. In Eq.~(\ref{58}), for an non-dividend American put option, $V^a(S(u),u)=\Phi(S(u))$ means $(S(u),u)$ is in the stopping region and $H(s(u),u) = rK$ \cite{peskir2006optimal}. Thus, we have the following  expression for the premium,
\begin{equation*}
\begin{aligned}
    p(s, t) &= \mathbf{E}_{t, s}\left[\int_t^TrKe^{-r(u-t)}\textit{I}\{s(u)\leq S^*(u)\}du\right] \\ &= rK\int_t^Te^{-r(u-t)}N\big(-\frac{1}{\sigma \sqrt{u-t}} [\ln \frac{s}{S^*(u)} + (r - \frac{\sigma^2}{2})(u-t)]\big)du \\ &= rK\int_t^Te^{-r(u-t)}N(-d_2(s, u-t, S^*(u)))du, 
\end{aligned}  
\end{equation*}
Here $d_2$ is defined in Eq.~(\ref{13}). Therefore, the value of an American put option can be written as 
\begin{equation*}
    V^a(s, t) = V^e(s, t) + rK\int_t^Te^{-r(u-t)}N(-d_2(s, u-t, S^*(u)))du.  
\end{equation*}
The unknown boundary $S^*(t)$ satisfies the following nonlinear Volterra integral equation
\begin{align*}
    K - S^*(t) &= V^e(S^*(t), t) + rK\int_t^Te^{-r(u-t)}N\big(-\frac{1}{\sigma \sqrt{u-t}} [\ln \frac{S^*(t)}{S^*(u)} + (r - \frac{\sigma^2}{2})(u-t)]\big)du  \notag\\ &= V^e(S^*(t), t) + rK\int_t^Te^{-r(u-t)}N\big(-d_2(S^*(t), u-t, S^*(u))\big)du. 
\end{align*}
However, $S^*(t)$ has no explicit formula and is challenging to determine, which makes it difficult to find $p(s, t)$. As a result, the value function $V^a(s, t)$ lacks a closed-form analytical solution.

The properties of $S^*(t)$ have been widely studied in the literature, with a key property being its continuity in $t$ \cite{peskir2006optimal}. From the integral representation, $p(s, t)$ is differentiable when $t < T$ and vanishes at $t = T$. However, $V^e(s,t)$ exhibits two types of singularities. First, it is non-differentiable at $s=K$ when $t=T$. Second, it contains $\sqrt{\tau}$ in denominators, which approaches $0$ as $t \rightarrow T$.  As a result, the term $p (s, t)$ is relatively easy for neural networks to approximate, whereas $V^e(s,t)$ is challenging due to these singularities. Given that European options share the same terminal conditions as American options, the term $V^e(s, t)$ not only encapsulates the singularity of $V^a(s,t)$ but also adheres to the terminal conditions.

To solve for $V^a(s,t)$, it is crucial to design a function $g_2$ that satisfies two key properties. First, it exactly satisfies the terminal conditions. Second, it incorporates both types of singularities inherent in  $V^e(s,t)$. Such design makes the residual component $V^a(s,t)-g_2(s,t)$ smoother and easier for the neural network to approximate.
A straightforward approach would be to set $g_2(s,t)=V^e(s,t)$. However, directly calculating $V^e(s,t)$ is computationally expensive. Instead, $g_2(s, t)$ is constructed as approximations of $V^e(s,t)$ that retain the singularities and exactly satisfy the terminal conditions, providing a more efficient and practical alternative.

Consider the term \begin{equation*}
    \Tilde{d_0}(s, \tau, K) = \frac{1}{2}\left[\Tilde{d_1}(s, \tau, K) + \Tilde{d_2}(s, \tau, K)\right] = -\frac{1}{\sigma \sqrt{\tau}}(\ln \frac{s}{K} + r\tau).    
\end{equation*}
The Taylor expansion of $N(x)$ at $\Tilde{d_0}$ can be written as 
\begin{equation*}
    N(x) = N(\Tilde{d_0}) + N'(\Tilde{d_0})(x-\Tilde{d_0}) + {{\scriptscriptstyle\mathcal{O}}}(x-\Tilde{d_0}).  
\end{equation*}
Therefore $N(\Tilde{d_1})$ and $N(\Tilde{d_2})$ can be expanded as 
\begin{align*}
    N(\Tilde{d_1}) &= N(\Tilde{d_0}) + N'(\Tilde{d_0})(\Tilde{d_1}-\Tilde{d_0}) + {{\scriptscriptstyle\mathcal{O}}}(\Tilde{d_1}-\Tilde{d_0}) = N(\Tilde{d_0}) - \frac{\sigma}{2}\sqrt{\tau} N'(\Tilde{d_0})+ {{\scriptscriptstyle\mathcal{O}}}(\sqrt{\tau}), \\ N(\Tilde{d_2}) &= N(\Tilde{d_0}) + N'(\Tilde{d_0})(\Tilde{d_2}-\Tilde{d_0}) + {{\scriptscriptstyle\mathcal{O}}}(\Tilde{d_2}-\Tilde{d_0})= N(\Tilde{d_0}) + \frac{\sigma}{2}\sqrt{\tau} N'(\Tilde{d_0})+ {{\scriptscriptstyle\mathcal{O}}}(\sqrt{\tau}).
\end{align*}
Then the value of European put can be written as \begin{equation*}
\begin{aligned}
V^e(s, t) &= Ke^{-r\tau}\cdot N\big(\Tilde{d_2}(s, \tau, K)\big) - s\cdot N\big(\Tilde{d_1}(s, \tau, K)\big) \\ 
&=  Ke^{-r\tau}\cdot [N(\Tilde{d_0}) + \frac{\sigma}{2}\sqrt{\tau} N'(\Tilde{d_0})+ {{\scriptscriptstyle\mathcal{O}}}(\sqrt{\tau})] - s\cdot [N(\Tilde{d_0}) - \frac{\sigma}{2}\sqrt{\tau} N'(\Tilde{d_0})+ {{\scriptscriptstyle\mathcal{O}}}(\sqrt{\tau})]  \\ 
&= N(\Tilde{d_0}) \cdot (Ke^{-r\tau} - s) + \frac{\sigma}{2}\sqrt{\tau}N'(\Tilde{d_0}) \cdot (Ke^{-r\tau} + s) + {{\scriptscriptstyle\mathcal{O}}}(\sqrt{\tau}) \\ &=N(\Tilde{d_0}) \cdot (Ke^{-r\tau} - s) + \frac{\sigma \sqrt{\tau}}{2\sqrt{2\pi}}e^{-\frac{\Tilde{d_0}^2}{2}} \cdot (Ke^{-r\tau} + s) + {{\scriptscriptstyle\mathcal{O}}}(\sqrt{\tau}).   
\end{aligned}
\end{equation*}
Denote the first and second terms of the Taylor expansion of $V^e(s,t)$ as
 \begin{equation*}
   V_1^e(s, t) = N(\Tilde{d_0}) \cdot (Ke^{-r\tau} - s), \quad
V_2^e(s, t) = \frac{\sigma \sqrt{\tau}}{2\sqrt{2\pi}}e^{-\frac{\Tilde{d_0}^2}{2}} \cdot (Ke^{-r\tau} + s), 
\end{equation*}
The value of $V^e(s,t)$ is then expressed as
\begin{equation*}
V^e(s, t) = V_1^e(s, t) + V_2^e(s, t) + {{\scriptscriptstyle\mathcal{O}}}(\sqrt{\tau}). 
\end{equation*} 
It is easy to prove that both $V_1^e(s, t)$ and $V_1^e(s, t) + V_2^e(s, t)$ satisfy the terminal condition (\ref{66}).

Based on these observations, three candidate functions can be considered for the role of $g_2$ in the exact terminal method, $V^e_1(s, t), V^e_1(s, t) + V^e_2(s, t)$, and $V^e(s, t)$ itself. Both $V^e_1$ and $ V^e_1 + V^e_2$
preserve the non-differentiability at the terminal $t=T$. 
However, only $V_1^e+V_2^e$
captures the singular behavior associated with the denominator containing $\sqrt{\tau}$ as $t\rightarrow T$. Consider the behavior of these functions as $s=K$ when $t\rightarrow T$. The option when $s=K$ is called an at-the-money option.
It is widely recognized in the literature that at-the-money options nearing their expiration date tend to approximate the value of $\frac{1}{\sqrt{2\pi}}s\sigma \sqrt{\tau}$ \cite{brenner1988simple}. For $V_1^e$ and $V_1^e+V_2^e$, consider the behavior at $s = K$  when $t \to T$, and ignore higher-order terms than $\sqrt{T-t} = \sqrt{\tau}$,
\begin{align*}
    \lim \limits_{t \to T}V_1^e(s, t) &= N(0)\cdot K (e^{-r\cdot 0} - 1) = 0,    \notag\\ \lim \limits_{t\to T}V_2^e(s, t) &= \frac{\sigma \sqrt{\tau}}{2\sqrt{2\pi}}e^{-\frac{r^2 \cdot 0}{2\sigma^2}} \cdot K (e^{-r\cdot 0} + 1) = \frac{1}{\sqrt{2\pi}} K \sigma \sqrt{\tau}. 
\end{align*}
These results indicate that the behavior of $V^e_1(s, t) + V^e_2(s, t)$ captures the singular features present in $V^e(s, t)$ at $s=K$ nearing $t=T$, while $V^e_1(s, t)$ alone does not. Therefore, the residual part of $V^a$ subtracting $V^e_1$ alone does not prevent the network from encountering challenges associated with learning the singular features.

In summary, $V_1^e(s, t)$ satisfies the terminal conditions, while $V_2^e(s, t)$  captures the singular behavior of the option value around $s=K$ as $t$ approaches the expiration $T$. The combination $V_1^e(s, t) + V_2^e(s, t)$ simultaneously satisfies the terminal conditions and matches the singularity of $V^e(s, t)$. Moreover, it has a lower computational cost than $V^e(s, t)$. Therefore, $V_1^e(s, t) + V_2^e(s, t)$ is a suitable candidate for the exact terminal function $g_2(s, t)$. In the following analysis, we evaluate the performance of our ETCNN when $g_2$ takes these three functions $V_1^e, V_1^e+V_2^e$, and $V^e$ separately to determine which function yields the best results through a comparative analysis.

\begin{figure}[!h]
  \subfloat[Total loss $\mathcal{L}$ for models with input normalization]{\includegraphics[width=0.47\textwidth]{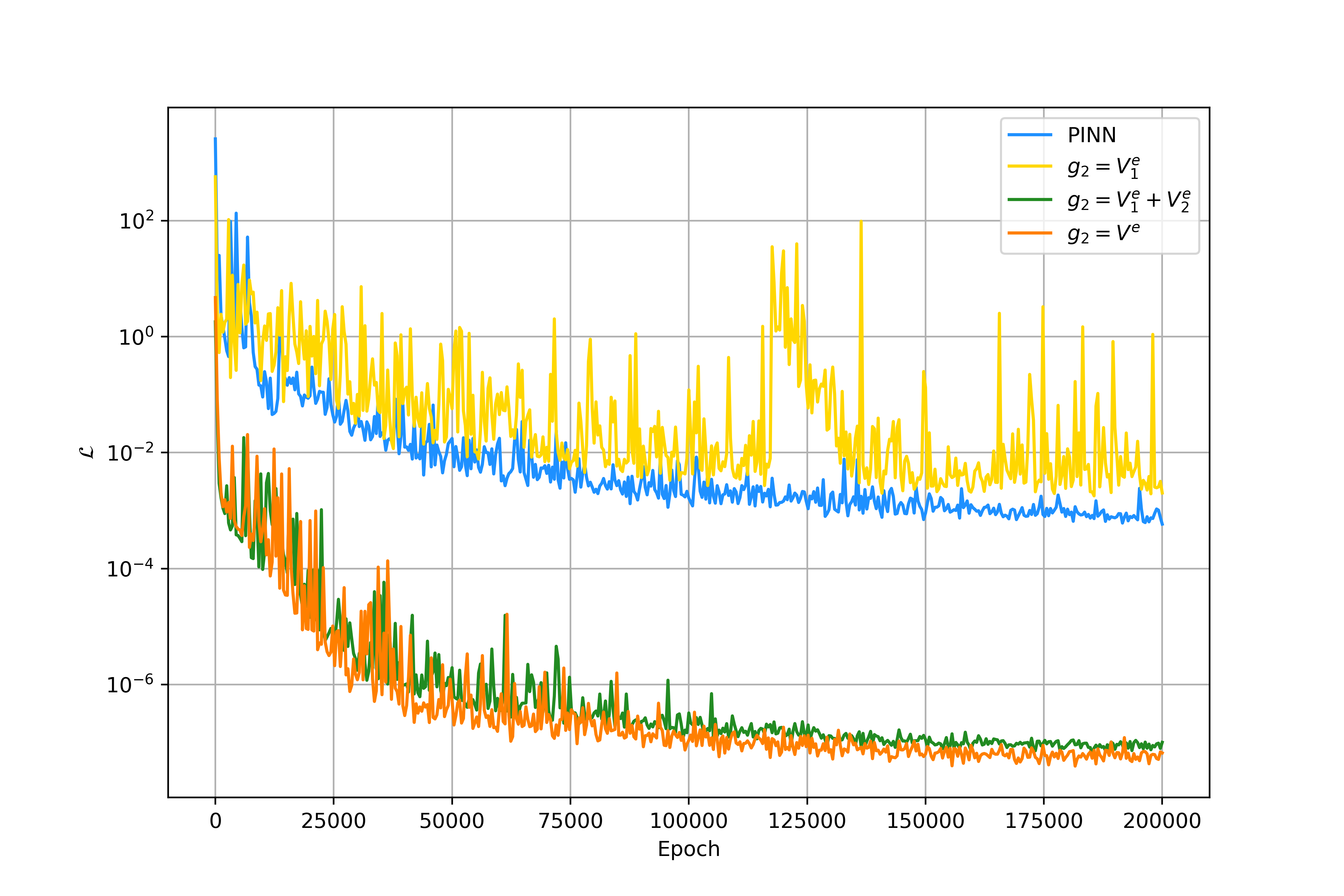}}
 \hfill 	
  \subfloat[$\mathcal{L}_{bs}$ for models with input normalization]{\includegraphics[width=0.47\textwidth]{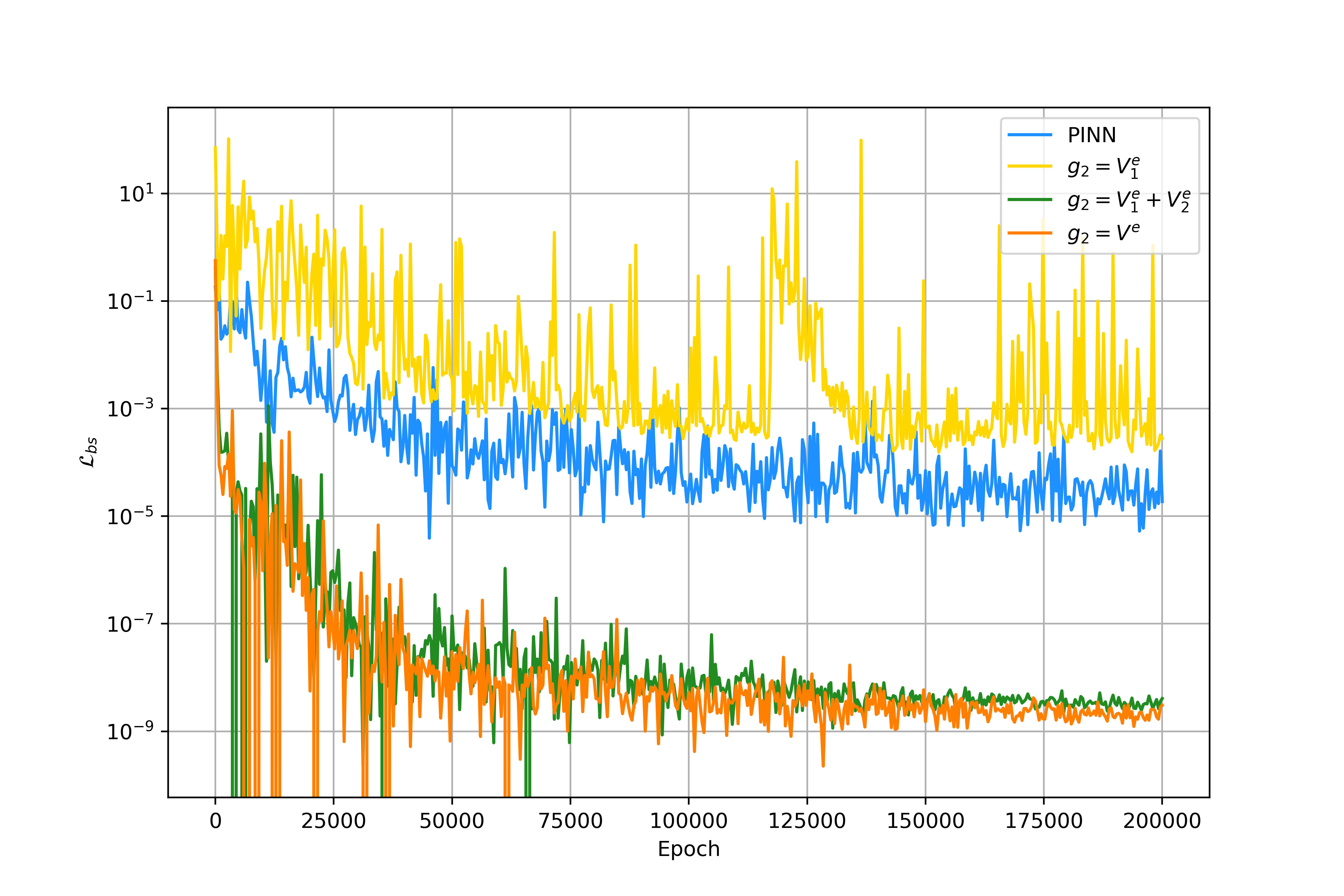}}
  \newline
    \subfloat[$\mathcal{L}_{tv}$ for models with input normalization]{\includegraphics[width=0.47\textwidth]{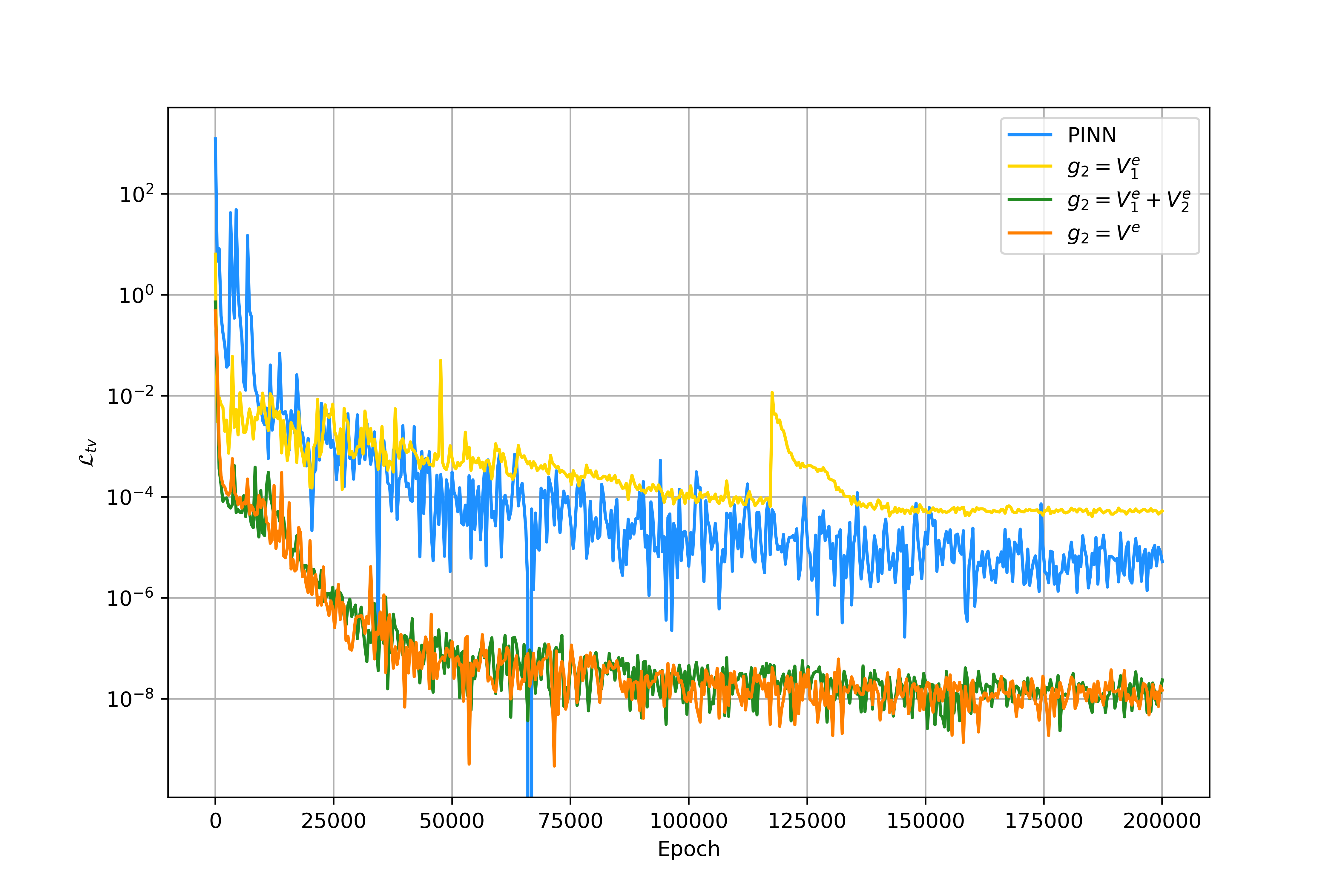}} 
 \hfill 	
  \subfloat[$\mathcal{L}_{eq}$ for models with input normalization]{\includegraphics[width=0.47\textwidth]{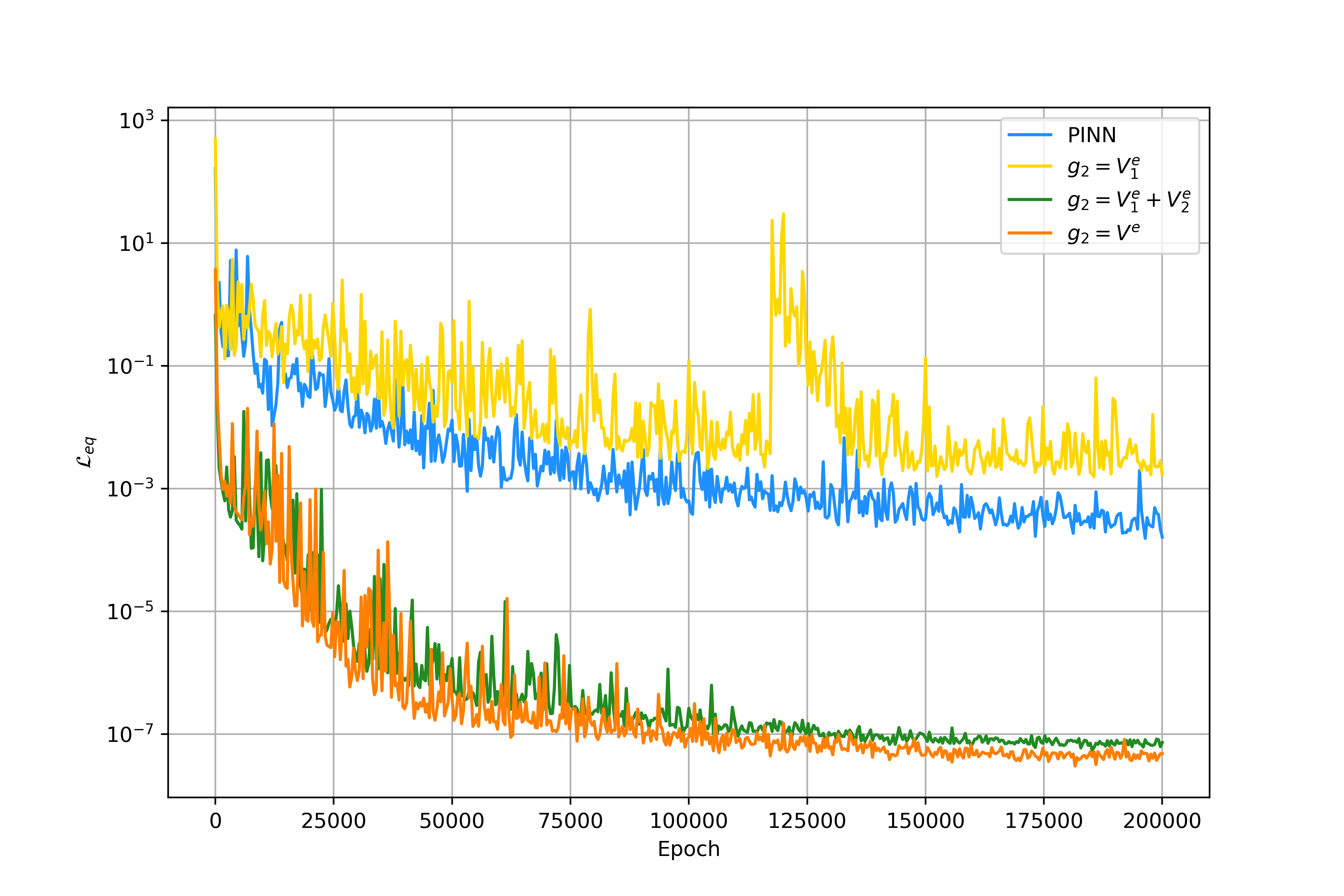}}
\caption{Total loss and each loss terms for the four models.}
    \label{fig8}
\end{figure}

\subsubsection{Numerical Results}

The parameters for the experiments are set as $K = 100, r = 0.02, T = 1, \sigma = 0.25$. To evaluate the performance of PINN and ETCNN, experiments are conducted using all three exact terminal functions as $g_2$ to compare their effectiveness. We define $g_1(s, t) = T-t$. The influence of input normalization on model performance is also examined. The binomial tree method with $N=4000$ steps is employed to compute the reference solution, serving as the benchmark for accuracy evaluation. We take $N_{tc} = 512$ sampling points to calculate $\mathcal{L}_{tc}$, and $N_{bs} = N_{tv} = N_{eq} = 4N_{tc}$ to calculate the other three loss terms. Training is performed over the interval $[20, 160] \times [0, T]$, with error evaluation focused on the subset $[60, 120] \times [0, T]$.

We first analyze the evolution of the loss terms during the training process. The total loss, along with each individual loss term for the four models with input normalization, are illustrated in Figure \ref{fig8}. The last term $\mathcal{L}_{tc} = 0$ is inherently satisfied in the exact terminal methods. Therefore we only present results for $\mathcal{L}_{bs}, \mathcal{L}_{tv}$ and $\mathcal{L}_{eq}$. All plots are shown in logarithmic scale. 
The results reveal that employing the exact terminal function with either $g_2(s, t) = V_1^e(s, t) + V_2^e(s, t)$ or $g_2(s, t) = V^e(s, t)$ significantly reduces the magnitude of each loss term. These values are several orders of magnitude smaller than those observed with PINN, demonstrating the effectiveness of the exact terminal methods. Additionally, ETCNN exhibits much faster convergence, achieving an accuracy comparable to that of a PINN trained for 200,000 epochs within only 10,000 epochs.
However, the performance of ETCNN with $g_2(s, t) = V_1^e(s, t)$ is noticeably inferior, with loss terms exceeding those of the standard PINN. This result suggests that $g_2(s, t) = V_1^e(s, t)$ is less effective compared to the other two choices of $g_2$ and highlights the importance of selecting an appropriate exact terminal function to ensure optimal model performance.

To further evaluate the accuracy of the solutions produced by the models, we employ two metrics, relative $L^2$ error and maximum absolute error (MAE). The relative $L^2$ error is defined as 
\begin{equation*}
    \varepsilon_{L^2} = \frac{||V^{true}-V^{pred}||_2}{||V^{true}||_2} = \frac{\sqrt{\frac{1}{N} \sum \limits_{i=1}^N (V_i^{true}-V_i^{pred})^2}}{\sqrt{\frac{1}{N}\sum \limits_{i=1}^N (V_i^{true})^2}},     
\end{equation*}
where $||\cdot||_2$ is the $L_2$ norm.  $V^{pred}(s, t)$ is solution given by the model. $N$ is the number of samples taken to calculate the error. The MAE is defined as follows, 
\begin{equation*}
    MAE = \max \limits_i |V_i^{true} - V_i^{pred}|.   
\end{equation*}
Each experiment was repeated three times, and the average results were recorded in Table \ref{table3}.
The result indicates that the accuracy of solutions obtained by models with input normalization is better than those without it. This improvement shows the importance of designing an input normalization layer to preserve the homogeneous structure of the option value function with respect to the asset price $s$.

\begin{table}[!htb]
\centering
\renewcommand{\arraystretch}{1.5}
\caption{Relative $L^2$ error and MAE in the ablation experiments. The first column on the left is the result of PINN. The last three columns are the results of ETCNN with three different exact terminal functions. The first two rows are the results of direct input, while the last two rows are the results of networks with input normalization. Bold font represents the best result in each row.}
\begin{tabular}{cc|cccc}
	\hline
	~ & Error & PINN & \makecell{ETCNN \\ $g_2 = V_1^e $} & \makecell{ETCNN \\ $g_2 = V_1^e+ V_2^e$} & \makecell{ETCNN \\ $g_2 = V^e$} \\ \hline
	
	\multirow{2}{*}{Without normalization} & Rel. $L^2$ error & $1.87\times 10^{-3}$ & $4.12\times 10^{-3}$  & $1.08\times 10^{-4}$ & $\textbf{6.91}\times \textbf{10}^{\textbf{-5}}$\\ 
	
	&MAE & $2.23\times 10^{-1}$  & $1.76\times 10^{-1}$ & $\textbf{8.52}\times \textbf{10}^{\textbf{-3}}$ & $1.06\times 10^{-2}$\\ \hline
	
	\multirow{2}{*}{With normalization} & Rel. $L^2$ error & $1.20\times 10^{-3}$ & $3.35\times 10^{-3}$ & $5.72\times 10^{-5}$ & $\textbf{5.34}\times \textbf{10}^{\textbf{-5}}$\\ 
	 
	& MAE & $1.16\times 10^{-1}$  & $1.31\times 10^{-1}$ & $\textbf{5.71}\times \textbf{10}^{\textbf{-3}}$ & $6.90\times 10^{-3}$\\ \hline
\end{tabular}
\label{table3}
\end{table}

ETCNN with both $g_2(s, t) = V^e(s, t)$ and $g_2(s, t) = V_1^e(s, t) + V_2^e(s, t)$ 
exhibit significant improvements in accuracy compared to PINN, reducing errors by 1-2 orders of magnitude. In terms of the $L^2$ error, ETCNN with $g_2 = V^e$ achieves the lowest error, followed closely by $g_2 = V_1^e+V_2^e$. In terms of MAE, the lowest value is obtained by ETCNN with $g_2 = V_1^e+V_2^e$, followed by $g_2 = V^e$. When input normalization is applied, the performance of ETCNN with $g_2 = V_1^e+V_2^e$ is nearly indistinguishable from ETCNN with $g_2 = V^e$, both achieving remarkably low error values. Specifically, the $L^2$ errors for both models are approximately $5\times 10^{-5}$, while their MAE values are around $6\times 10^{-3}$. These results imply that, for an option with an exercise price of $K=100$ units of currency, the maximum absolute error of the solutions obtained using ETCNN with $g_2 = V_1^e+V_2^e$ or $g_2 = V^e$ is less than one cent, which is a very low error in practice. 
In contrast, ETCNN with $g_2 = V_1^e$ yields the poorest performance, even worse than PINN. This outcome highlights that not all solutions that exactly satisfy boundary conditions lead to lower errors. 
Merely fulfilling the terminal conditions is insufficient to ensure accurate solutions. The effectiveness of the method is strongly dependent on the smoothness and structure of the residual difference between $V^{true}$ and the chosen $g_2$. This observation emphasizes the critical importance of selecting an appropriate $g_2$.

\begin{figure}[!h]
    \centering
\includegraphics[width=0.99\textwidth]{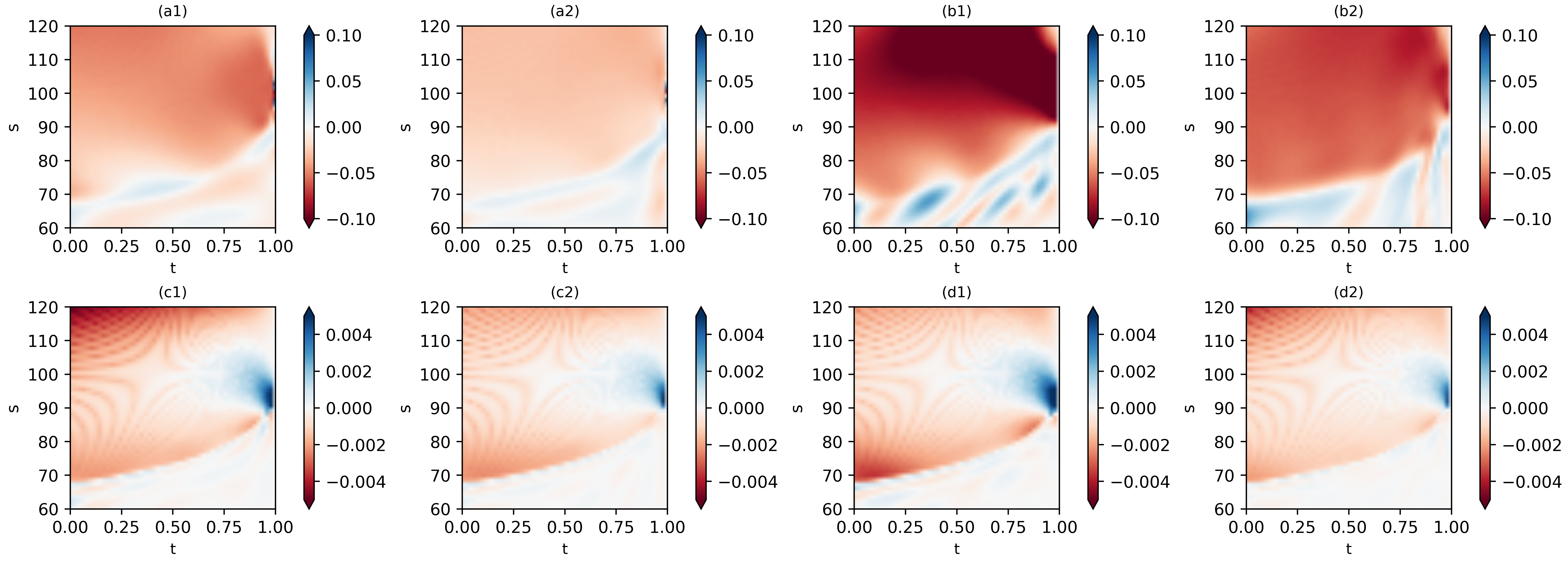}\\
\caption{Pointwise difference between the true solution and predicted solution obtained by the PINN and ETCNN with three $g_2$ functions. $(a1)$, PINN. $(a2)$, PINN + Norm. $(b1)$, ETCNN $(g_2 = V_1^e)$. $(b2)$, ETCNN $(g_2 = V_1^e)$  + Norm. $(c1)$, ETCNN $(g_2 = V_1^e+V_2^e)$. $(c2)$, ETCNN $(g_2 = V_1^e+V_2^e)$  + Norm. $(d1)$, ETCNN $(g_2 = V^e)$. $(d2)$, ETCNN $(g_2 = V^e)$ + Norm.}
    \label{fig9}
\end{figure}

To better illustrate the results of the previous table, Figure \ref{fig9} shows the pointwise difference between the predicted solution by PINN or ETCNN and the true solution. The pointwise difference is defined as the difference between the true solution and the model prediction, \begin{equation*}
    \text{Diff}(s, t) = V^{true}(s, t) - V^{pred}(s, t).    
\end{equation*}
Models incorporating an input normalization layer consistently achieve lower absolute errors compared to those without. Among the ETCNN models, $g_2 = V^e$ and $g_2 = V_1^e+V_2^e$ exhibit far superior performance relative to PINN, whereas the performance of $g_2 = V^e_1$ falls below that of PINN. 
Notably, when input normalization is applied, 
the performance of $g_2 = V_1^e + V_2^e$ is comparable to that of $g_2 = V^e$.
However, computing $V^e$ is more computationally expensive than calculating $V_1^e + V_2^e$, as it requires evaluating the cumulative distribution function twice at different points. Therefore, we select the ETCNN model using $g_2 = V_1^e + V_2^e$ and with an input normalization layer as the final model. A detailed analysis of this model's results is provided in the following paragraph.

\begin{figure}[!h]
    \centering
\includegraphics[width=0.9\textwidth]{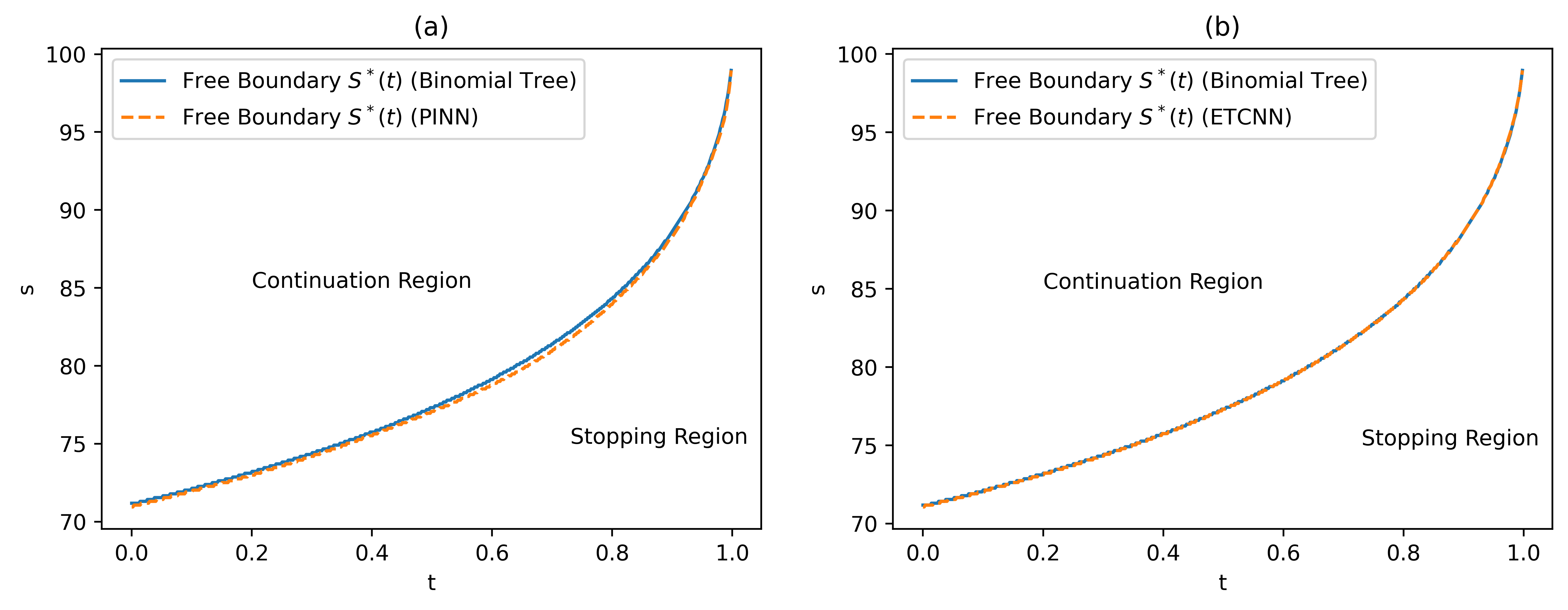}\\
\caption{Early exercise boundary. (a), Early exercise boundary obtained by the exact solution and PINN. (b), Early exercise boundary obtained by the exact solution and ECTNN with $g_2(s, t) = V^e_1(s, t) + V^e_2(s, t)$. Both PINN and ETCNN implement input normalization.}
        \label{fig12}
\end{figure}

Additionally, Figure \ref{fig9} reveals an increase in error near the early exercise boundary, highlighting the added complexity of solving equations with unknown free boundaries. This behavior reflects the additional challenge introduced by the free-boundary feature in option pricing. Figure \ref{fig12} compares the free boundary obtained from the true solution, the PINN solution, and the ETCNN solution with $g_2(s, t) = V^e_1(s, t) +V^e_2(s, t)$.
The early exercise boundary derived from ETCNN closely aligns with the true early exercise boundary, showing near-complete overlap. In contrast, the early exercise boundary derived from the PINN shows noticeable deviations, particularly at points far from the expiration date. 
These results show the capability of our method to effectively handle the challenges posed by free boundary problems.

We extend our analysis by comparing the performance of ETCNN with some traditional methods: Barone-Adesi and Whaley (BAW) method, the binomial tree (BT), finite difference (FD), and least squares Monte Carlo (LSM) method. For the BT and FD methods, results are evaluated using $N = 100, 200$, and $400$ steps. In the case of the LSM approach, simulations are performed with $M = 10,000$ paths, and results are obtained for $N = 100, 200$, and $400$ steps as well. The comparative results of these methods are summarized in Table \ref{table4}.

\begin{table}[!ht]
    \centering
    \caption{Relative $L^2$ error and MAE comparison of ETCNN, PINN, and traditional numerical methods. ETCNN takes $g_2 = V_1^e+V_2^e$. Both PINN and ECTNN implement input normalization. Bold font represents the best result.}
 \label{table4}
\renewcommand{\arraystretch}{1.5} 
    \begin{tabular}{ccc|ccc}
    \hline
        Methods & Rel. $L^2$ error & MAE & Methods & Rel. $L^2$ error & MAE\\ \hline
        ETCNN & $\textbf{5.72} \times \textbf{10}^{\textbf{-5}}$ &$\textbf{5.71}\times \textbf{10}^{\textbf{-3}}$ & FD($N = 100$) & $5.36\times 10^{-4}$ & $3.82\times 10^{-2}$ \\ 
        PINN & $1.20\times 10^{-3}$ & $1.16\times 10^{-1}$ & FD($N = 200$)& $2.78\times 10^{-4}$ & $2.00\times 10^{-2}$\\ 
        BAW
          & $1.49 \times 10^{-3}$  & $9.03 \times 10^{-2}$ & FD($N = 400$)& $1.46\times 10^{-4}$ & $1.05\times 10^{-2}$\\
        BT($N = 100$) & $3.99\times 10^{-4}$ & $2.46\times 10^{-2}$& LSM($N = 100$)& $2.38\times 10^{-3}$ & $2.24\times 10^{-1}$\\
        BT($N = 200$) & $1.97\times 10^{-4}$ & $1.27\times 10^{-2}$& LSM($N = 200$)& $2.45\times 10^{-3}$ & $2.53\times 10^{-1}$\\ BT($N = 400$) & $9.78\times 10^{-5}$ & $6.61\times 10^{-3}$ & LSM($N = 400$)& $2.61\times 10^{-3}$ & $2.21\times 10^{-1}$\\
        \hline
    \end{tabular}
\end{table}

The performance of the LSM method is the least favorable among the evaluated approaches, while the BAW approximation performs slightly better but remains among the less accurate methods. PINN yields higher accuracy than both LSM and BAW but still falls short of the FD and BT methods. This suggests that solutions obtained using a neural network without any specific design are less accurate than traditional numerical methods. 
In contrast, our ETCNN not only surpasses PINN but also outperforms several traditional numerical methods in terms of accuracy. Specifically, our ETCNN achieves the lowest relative $L^2$ error and the smallest MAE. Among all evaluated methods, only ETCNN and the BT method with larger step sizes can achieve an MAE on the order of $10^{-3}$, which is a huge advantage in actual transactions.

To further illustrate the accuracy of the proposed approach, Table~\ref{table5} reports the option values at several fixed $(s, t)$ points ($t$ is the current time and $T-t$ is the time to maturity), along with the average computation time required by each method. For the ETCNN and PINN models, the reported times correspond to the evaluation phase of the trained networks, excluding the training process. The remaining model parameters are fixed as before ($K = 100$, $r = 0.02$, $\sigma = 0.25, T=1$). The reference values in the first column are obtained using the BT method with $N = 4000$, which serves as a benchmark for the true solution. The second column presents the corresponding European option prices under the same parameters, computed from Eq.~(\ref{16a}). The difference between these two columns represents the early exercise premium. The subsequent columns list the results obtained from the proposed ETCNN ($g_2 = V_1^e+V_2^e$), PINN, BAW, BT ($N = 400$), FD ($N = 400$), and LSM ($N = 400$) methods. This comparison clearly shows the deviations of each method from the benchmark solution.

\begin{table}[!ht]
\small
\centering
\caption{The price of American put options with fixed model parameters ($K = 100$, $r = 0.02$, $\sigma = 0.25, T=1$) calculated by ETCNN, PINN, and traditional methods. The binomial tree method with $N=4000$ serves as the reference solution. $V^e(s, t)$ denotes the price of European options with the same parameters (calculated by Eq.~(\ref{16a})). Bold represents the solution closest to the reference solution among the six methods. The average time consumption (in milliseconds) for each method represents the mean evaluation time over all sample points listed in the table.} 
\label{table5}
\renewcommand{\arraystretch}{1.3}   

\begin{tabular}[c]{cc|cccccccc}  	
\hline \makecell{Time to \\ expiration \\ $T-t$} & \makecell{Underlying \\ price $s$} &	\makecell{Ref. solution \\ BT(N=4000)} & \makecell{European \\ $V^e(s, t)$}   & ETCNN &PINN  & BAW  & \makecell{BT \\ (N=400)} & \makecell{FD \\ (N=400)} & \makecell{LSM \\ (N=400)} \\ \hline 

\multirow{5}{*}{0.25} & 80 & 20.010  & 19.684  & 20.011  & 20.022  & 20.000  & \textbf{20.010}  & 20.009  & 20.036  \\ 
&  90 & 11.030  & 10.914  & \textbf{11.030}  & 11.052  & 10.999  & 11.032  & 11.029  & 11.005 \\
& 100 & 4.759  & 4.726  & \textbf{4.759}  & 4.786  & 4.752  & 4.757  & 4.754  & 4.660    \\
& 110 & 1.571  & 1.563  & \textbf{1.572}  & 1.603  & \textbf{1.572}  & 1.573  & \textbf{1.570}  & 1.562  \\ 
& 120 & 0.403  & 0.401  & 0.404  & 0.436  & 0.404  & \textbf{0.403}  & \textbf{0.403} & 0.417
\\	\hline   

\multirow{5}{*}{0.5} & 80 & 20.306  & 19.875  & \textbf{20.307}  & 20.323  & 20.247  & \textbf{20.307}  & \textbf{20.305}  & 20.475 \\ 
&   90 & 12.289  & 12.101  & \textbf{12.289}  & 12.314  & 12.249  & 12.287  & 12.285  & 12.239     \\
&   100 & 6.597  & 6.522  & \textbf{6.598}  & 6.626  & 6.587  & 6.595  & 6.590  & 6.665       \\ 
&  110 & 3.155  & 3.127  & \textbf{3.156}  & 3.185  & 3.157  & 3.159  & 3.153  & 3.158        \\ 
& 120 & 1.361  & 1.351  & \textbf{1.362}  & 1.390  & 1.366  & \textbf{1.362}  & 1.359  & 1.333 \\ \hline 

\multirow{5}{*}{0.75} & 80 & 20.761  & 20.228  & \textbf{20.762}  & 20.778  & 20.684  & \textbf{20.762}  & 20.757  & 20.671 \\
& 90 & 13.336  & 13.075  & 13.337  & 13.360  & 13.290  & \textbf{13.336}  & 13.333  & 13.341  \\
& 100 & 7.955  & 7.832  & \textbf{7.956}  & 7.983  & 7.942  & 7.951  & 7.946  & 8.036  \\
& 110 & 4.435  & 4.379  & \textbf{4.436}  & 4.463  & 4.439  & 4.439  & 4.429  & 4.475 \\
& 120 & 2.332  & 2.307  & \textbf{2.333}  & 2.361  & 2.341  & 2.334  & \textbf{2.331}  & 2.317 \\ \hline 

\multicolumn{2}{c|}{\makecell{Ave. computational \\ time (ms)}}  & - & - & 1.095  & 0.447  & 7.654  & 60.22  & 453.98  & 476.11
\\ \hline 
\end{tabular}  
\end{table}

The results show that the proposed method consistently achieves the highest accuracy across various $(s, t)$ cases, including scenarios close to maturity as well as those far from maturity, and for options that are in-the-money, at-the-money, or out-of-the-money. This demonstrates the strong robustness and stability of the proposed approach under different market conditions. 
Furthermore, a trained ETCNN can efficiently compute option values at a large number of points, whereas traditional methods require longer computation time, especially when estimating values across multiple points. For instance, both the BT and FD methods require $N$ iterative steps, calculating each step sequentially. Similarly, LSM needs to generate $M$ simulation paths and perform least squares fitting. The BAW method, although faster than other numerical schemes, still cannot match the evaluation speed of a well-trained ETCNN when applied repeatedly. Therefore, our ETCNN with a carefully designed $g_2$ function not only offers superior accuracy but also provides an advantage in computational speed.

\subsection{American Call Options with Dividends}

\label{Sec4.2}
The valuation of a single-asset American call option without dividends is equivalent to its European counterpart, making its value straightforward to compute. However, when the underlying asset pays dividends, the American call option typically holds a premium over its European counterpart, and its value does not have a closed-form solution.

For an American call option on an asset with a continuous dividend yield $q$, its value satisfies the following modified
version linear complementarity conditions, 
\begin{numcases}{}
\quad  V(s, t) \geq (s-K)^+,\quad  \forall s \geq 0, t \in [0, T),    \notag \\ \quad  
\frac{\partial V}{\partial t} + \frac{1}{2}\sigma^2 s^2 \frac{\partial^2 V}{\partial s^2} + (r-q)s\frac{\partial V}{\partial s}-rV \leq 0,\quad  \forall  s \geq 0, t \in [0, T),   \notag \\ \quad \big(\frac{\partial V}{\partial t} + \frac{1}{2}\sigma^2 s^2 \frac{\partial^2 V}{\partial s^2} + (r-q)s\frac{\partial V}{\partial s}-rV  \big)\cdot \big(V-(s-K)^+ \big)=0,\quad  \forall  s \geq 0,  t \in [0, T),   \notag \\ \quad  V(s, T) = (s-K)^+.   \notag 
\end{numcases}
The exact solution for its European counterpart differs slightly from Eq.~(\ref{54}) due to the inclusion of the dividend term. The modified form is given as follows, 
\begin{equation}
    V(s, t) = se^{-q\tau}N(\hat{d_1}(s, \tau, K)) - Ke^{-r\tau}N(\hat{d_2}(s, \tau, K)).
\label{new_17a}
\end{equation}
Here, $\hat{d_1}, \hat{d_2}$ are defined as : \begin{align*} 
\left\{  
\begin{aligned} \quad &
   \hat{d_1}(s, \tau, K) = \frac{1}{\sigma \sqrt{\tau}}\big(\ln{\frac{s}{K}}+(r-q+\frac{\sigma^2}{2})\tau\big), \\ \quad &
   \hat{d_2}(s, \tau, K) = \hat{d_1}(s, \tau, K) - \sigma \sqrt{\tau} = \frac{1}{\sigma \sqrt{\tau}}\big(\ln{\frac{s}{K}}+(r-q-\frac{\sigma^2}{2})\tau\big). 
   \end{aligned}
\right.       
\end{align*}
Similar to the previous example, we take \begin{equation*}
    \hat{d_0}(s, \tau, K) = \frac{1}{2}(\hat{d_1}(s, \tau, K) + \hat{d_2}(s, \tau, K)).   
\end{equation*}

By applying a first-order Taylor expansion to $N(\cdot)$ and substituting the result into Eq.~$(\ref{new_17a})$, an exact terminal function $g_2(s, t)$ is derived for single-asset American call options with continuous dividend payments, \begin{equation}
    g_2(s, t) = N(\hat{d_0})[se^{-q\tau}-Ke^{-r\tau}] + \frac{1}{2\sqrt{2\pi}}\sigma \sqrt{\tau}e^{-\frac{\hat{d_0}^2}{2}}[se^{-q\tau} + Ke^{-r\tau}]. \label{103}
\end{equation}
The output of our ETCNN in this experiment is $\Tilde{u}_{NN}(s, t) = g_1(s, t)u_{NN}(s, t) + g_2(s, t)$, where $g_2$ is defined in Eq.~(\ref{103}) and $g_1(s,t) = T-t$.

In this example, we set $K=200, r = 0.05, T = 2$. To demonstrate the robustness of our ETCNN, various parameter configurations have been applied.
The volatility level $\sigma$ is chosen to take values of $0.1, 0.25$, and $0.4$, representing markets with different volatility conditions. Dividend yields are varied across $\{0.01, 0.03, 0.05, 0.07\}$ to assess applicability to both low-dividend and high-dividend-paying assets.
Input normalization is applied in all experiments for both PINN and ETCNN models. 
The number of sampling points for calculating $\mathcal{L}_{tc}$ is $N_{tc} = 8192$, while $\mathcal{L}_{bs}, \mathcal{L}_{tv}$ and $\mathcal{L}_{eq}$ uses $4N_{tc}$ points. Training is conducted over 
$[0, 800] \times [0, T]$, with evaluation of errors on the subdomain $[160, 240]\times [0, T]$ to measure performance. The solution obtained by the binomial tree method with $N=4000$ serves as the reference for the true solution. The relative $ L^2$ errors between the solutions obtained from different methods and the reference solution are compared. The results are reported in Table \ref{table6}.

\begin{table}[!ht]
    \centering
    \caption{Relative $L^2$ error for American call options with different volatility $\sigma$ and dividend rate $q$. Comparisons are made between our ETCNN and other methods. Both PINN and ETCNN apply the input normalization technique. Bold font represents the best result in each column.}
 \label{table6}
\renewcommand{\arraystretch}{1.5} 
    \begin{tabular}{c|cccc}
    \hline \multirow{2}{*}{Methods} & \multicolumn{4}{c}{$\sigma = 0.1$}   \\
          & $q = 0.01$  & $q = 0.03$ & $q = 0.05$ & $q = 0.07$ \\ \hline
        ETCNN & $3.79\times 10^{-5}$& $\textbf{5.98}\times \textbf{10}^{\textbf{-5}}$& $1.69\times 10^{-4}$& $4.15\times 10^{-4}$\\ 
        PINN & $2.81\times 10^{-3}$ & $3.00\times 10^{-3}$& $4.48\times 10^{-3}$& $4.02\times 10^{-3}$\\
        BAW & $\textbf{1.05}\times \textbf{10}^{\textbf{-5}}$ & $4.02\times 10^{-4}$ & $1.86\times 10^{-3}$& $2.45\times 10^{-3}$\\
        BT(N = 400) & $9.34\times 10^{-5}$& $9.67\times 10^{-5}$& $\textbf{9.87}\times \textbf{10}^{\textbf{-5}}$& $\textbf{6.48}\times \textbf{10}^{\textbf{-5}}$\\ 
        FD(N = 400) & $1.06\times 10^{-4}$& $1.06\times 10^{-4}$& $1.68\times 10^{-4}$& $3.01\times 10^{-4}$\\
        LSM(N = 400) & $3.02\times 10^{-3}$& $5.27\times 10^{-3}$& $3.67\times 10^{-3}$& $2.53\times 10^{-3}$ \\
    \hline
        \multirow{2}{*}{Methods} & \multicolumn{4}{c}{$\sigma = 0.25$}   \\  & $q = 0.01$  & $q = 0.03$ & $q = 0.05$ & $q = 0.07$\\ \hline ETCNN & $7.16\times 10^{-5}$& $\textbf{1.78}\times \textbf{10}^{\textbf{-4}}$& $\textbf{2.18}\times \textbf{10}^{\textbf{-4}}$& $\textbf{2.23}\times \textbf{10}^{\textbf{-4}}$\\
        PINN & $2.43\times 10^{-3}$ & $3.28\times 10^{-3}$& $3.86\times 10^{-3}$& $4.33\times 10^{-3}$\\ BAW & $\textbf{5.15}\times \textbf{10}^{\textbf{-5}}$ & $2.98\times 10^{-3}$ & $4.49\times 10^{-3}$& $5.12\times 10^{-3}$\\
        BT(N = 400) & $2.33\times 10^{-4}$& $2.59\times 10^{-4}$& $2.72\times 10^{-4}$& $2.52\times 10^{-4}$\\ 
        FD(N = 400) & $1.77\times 10^{-4}$& $2.12\times 10^{-4}$& $3.59\times 10^{-4}$& $5.74\times 10^{-4}$\\
        LSM(N = 400) & $1.45\times 10^{-2}$& $1.06\times 10^{-2}$& $8.27\times 10^{-3}$& $7.52\times 10^{-3}$ \\
    \hline \multirow{2}{*}{Methods} & \multicolumn{4}{c}{$\sigma = 0.4$}   \\
    & $q = 0.01$ & $q = 0.03$ & $q = 0.05$ & $q = 0.07$\\ \hline ETCNN & $\textbf{8.45}\times \textbf{10}^{\textbf{-5}}$& $\textbf{1.85}\times \textbf{10}^{\textbf{-4}}$& $\textbf{1.68}\times \textbf{10}^{\textbf{-4}}$& $\textbf{2.53}\times \textbf{10}^{\textbf{-4}}$\\
        PINN & $2.57\times 10^{-3}$ & $2.72\times 10^{-3}$& $3.19\times 10^{-3}$& $3.72\times 10^{-3}$\\ BAW & $3.50\times 10^{-4}$ & $4.08\times 10^{-3}$ & $5.53\times 10^{-3}$ & $6.54\times 10^{-3}$\\
        BT(N = 400) & $2.96\times 10^{-4}$& $3.14\times 10^{-4}$& $3.24\times 10^{-4}$& $3.17\times 10^{-4}$\\  
        FD(N = 400) & $2.58\times 10^{-4}$& $3.24\times 10^{-4}$& $5.02\times 10^{-4}$& $7.01\times 10^{-4}$\\
        LSM(N = 400) & $1.81\times 10^{-2}$& $1.45\times 10^{-2}$& $1.11\times 10^{-2}$& $1.07\times 10^{-2}$ \\
    \hline  
    \end{tabular}
\end{table}

The results demonstrate that the proposed ETCNN outperforms other methods across most of the various scenarios, particularly in high-volatility markets. Compared to PINN, ETCNN achieves a reduction in $L^2$ error by an order of magnitude. 
When compared with traditional numerical methods, ETCNN exhibits superior robustness under challenging market conditions. As volatility and dividend yields increase, the errors associated with traditional methods grow substantially, whereas ETCNN maintains an $L^2$ error of approximately $2\times 10^{-4}$, showing its robustness and adaptability to diverse and fluctuating market conditions.

Among the tested approaches, LSM shows the poorest performance, while BAW performs slightly better than LSM and is comparable to PINN. It is worth noting that BAW achieves relatively high accuracy when the dividend yield is small, as the option price becomes closer to that of a European option under such conditions.
Overall, ETCNN outperforms traditional approaches and standard PINN in terms of accuracy across most scenarios.
This makes ETCNN a more reliable and practical solution compared to both traditional methods and standard PINN.
  
We further evaluated the proposed method under fixed option parameters by setting $K = 2$, $r = 0.05$, and $T = 2$, while varying the volatility, dividend yield, time, and underlying asset price. Table \ref{table5a} shows that our method consistently achieves the highest accuracy across different parameter settings, demonstrating strong robustness to changes in market conditions. Moreover, the trained network exhibits a high computational efficiency, achieving rapid inference once training is completed.

\begin{table}[!ht] 
\small
\centering
 \caption{The price of American call options with fixed model parameters ($K = 200$, $r = 0.05$, $T=2$) calculated by ETCNN, PINN, and traditional methods. The binomial tree method with $N=4000$ serves as the reference solution. $V^e(s, t)$ denotes the price of European options with the same parameters (calculated by Eq.~(\ref{new_17a})). Bold represents the solution closest to the reference solution among the six methods. The average time consumption (in milliseconds) for each method represents the mean evaluation time over all sample points listed in the table.} 
 \label{table5a}
\renewcommand{\arraystretch}{1.5}   
\begin{tabular}[c]{cc|cccccccc}  
 \hline \makecell{Option \\ parameters}&	\makecell{Underlying \\ price $s$} &	\makecell{Ref. solution \\ BT(N=4000)} & \makecell{European \\ $V^e(s, t)$}  & ETCNN & PINN & BAW  & \makecell{BT \\ (N=400)} & \makecell{FD \\ (N=400)} & \makecell{LSM \\ (N=400)}
\\ \hline \multirow{3}{*}{\makecell{$\sigma=0.25$ \\ $q=0.05$ \\ $T-t = 0.5$}} & 180 & 5.560  & 5.542  & \textbf{5.561}  & 5.650  & 5.577  & 5.555  & 5.555  & 5.715   \\  & 200 & 13.805  & 13.739  & \textbf{13.805}  & 13.890  & 13.824  & 13.798  & 13.792  & 13.755  \\  & 220 & 26.409  & 26.219  & \textbf{26.410}  & 26.484  & 26.413  & 26.416  & 26.405  & 26.149 
\\ \hline \multirow{3}{*}{\makecell{$\sigma=0.4$ \\ $q=0.05$ \\ $T-t = 0.5$}} & 180 & 12.553  & 12.506  & \textbf{12.556}  & 12.661  & 12.582  & 12.560  & 12.548  & 12.752   \\  & 200 & 22.044  & 21.937  & \textbf{22.047}  & 22.154  & 22.075  & 22.033  & 22.022  & 21.612 \\  & 220 & 34.288  & 34.073  & \textbf{34.288}  & 34.394  & 34.306  & 34.299  & 34.273  & 33.995 \\  \hline \multirow{3}{*}{\makecell{$\sigma=0.25$ \\ $q=0.07$ \\ $T-t = 0.5$}} & 180 & 5.095  & 5.025  & \textbf{5.098}  & 5.178  & 5.125  & 5.091  & 5.088  & 5.245   \\ & 200 & 12.966  & 12.722  & \textbf{12.969}  & 13.038  & 12.979  & 12.961  & 12.951  & 13.029  \\  & 220 & 25.321  & 24.670  & \textbf{25.326}  & 25.380  & 25.270  & 25.328  & 25.313  & 25.263\\
\hline \multirow{3}{*}{\makecell{$\sigma=0.25$ \\ $q=0.05$ \\ $T-t = 1$}} & 180 & 10.114  & 10.030  & 10.118  & 10.199  & 10.179  & \textbf{10.117}  & 10.104  & 10.388   \\  & 200 & 19.138  & 18.925  & \textbf{19.142}  & 19.221  & 19.213  & 19.129  & 19.120  & 19.300  \\  & 220 & 31.264  & 30.802  & 31.268  & 31.344  & 31.324  & \textbf{31.267}  & 31.260  & 30.956
\\	\hline   \multicolumn{2}{c|}{\makecell{Ave. computational \\ time (in ms)}}  & -  & - &   1.897  & 0.431  & 9.265  & 60.08  & 485.46  & 548.32
\\ \hline 
	\end{tabular}  
\end{table}

\subsection{Multi-Asset Options with Geometric Average Payoffs}
\label{Sec4.3}

In this example, the focus is on multi-asset American put options, where the payoff function is defined as the geometric mean of the prices of $n$ underlying assets.
Let $I = I(s_1, \cdots, s_n) = \big(\prod_{i=1}^{n} s_i(t)\big)^{\frac{1}{n}}$ be the geometric mean of $\{s_1, \cdots, s_n\}$. The payoff function is defined as
\begin{equation*}
    \Phi(s_1, \cdots, s_n) = \Big(K - \big(\prod_{i=1}^{n} s_i\big)^{\frac{1}{n}}\Big)^+ = (K-I)^+.    
\end{equation*}
The PDE operator is \begin{equation*}
\mathcal{F}(V(s_1, \cdots, s_n, t)) = \frac{\partial V}{\partial t} + \frac{1}{2} \sum \limits_{i,j=1}^n\sigma_i \sigma_j \rho_{ij}s_is_j \frac{\partial^2 V}{\partial s_i \partial s_j} + \sum \limits_{i=1}^n (r-q_i)s_i\frac{\partial V}{\partial s_i}-rV.        
\end{equation*}
Then the value of a multi-asset American put option on the geometric mean of $n$ underlying assets $V = V(s_1, \cdots, s_n)$ satisfies the following BSM equations,
\begin{numcases}{} \quad
V(s_1, \cdots, s_n, t) \geq (K-I)^+,\quad  \forall s_i \geq 0,  t \in [0, T),    \notag \\ \quad
\mathcal{F}(V(s_1, \cdots, s_n, t)) \leq 0,\quad  \forall s_i \geq 0, t \in [0, T),   \notag  \\ \quad \mathcal{F}(V(s_1, \cdots, s_n, t)) \cdot \big(V-(K-I)^+\big) = 0, \quad  \forall s_i \geq 0, t \in [0, T), \quad     \notag \\ \quad V(s_1, \cdots, s_n, T) = (K-I)^+, \quad \forall s_i \geq 0.   \notag 
\end{numcases}

The geometric average $I_t$ of $n$ geometric Brownian motion processes $\{S_i\}_{i=1}^n$ is itself a geometric Brownian motion \cite{eytan1986pricing}. From the high-dimensional It$\hat{\text{o}}$'s lemma, the dynamics of 
$I$ can be expressed as\begin{equation*}
    dI = \sum \limits_{i=1}^n \frac{\partial I}{\partial s_i}dS_i + \frac{1}{2}\sum \limits_{i, j = 1}^n\frac{\partial^2 I}{\partial s_i \partial s_j}dS_i dS_j.    
\end{equation*}
Substituting into the dynamics of $s_i$ in Eq.~(\ref{55}) and (\ref{56}) yields\begin{equation*}
        dI = \sum \limits_{i=1}^n \frac{\partial I}{\partial s_i}\left[(r-q_i)S_i + \sigma_i S_i dW_i\right] + \frac{1}{2}\sum \limits_{i, j = 1}^n\frac{\partial^2 I}{\partial s_i \partial s_j}\rho_{ij}\sigma_i \sigma_j dt .   
\end{equation*}
The first and second partial derivatives of 
$I$ are calculated as \begin{equation*}
    \frac{\partial I}{\partial s_i} = \frac{I}{nS_i}, \quad \frac{\partial^2 I}{\partial s_i \partial s_j} = \begin{cases}
\frac{1}{n^2}\frac{I}{S_iS_j}&\text{if}\ i \neq j,\\
\frac{1}{n}(\frac{1}{n}-1)\frac{I}{S_i^2}&\text{if}\ i=j.
\end{cases}   
\end{equation*}
Substituting these expressions into the dynamics of $I$, the simplified expression becomes\begin{equation*}
    dI(t) = \left[\frac{1}{n}\sum \limits_{i=1}^n(r-q_i - \frac{1}{2}\sigma_i^2) + \frac{1}{2n^2}\sum \limits_{i, j = 1}^n\rho_{ij}\sigma_i \sigma_j\right]I(t)dt + \frac{1}{n}\sum \limits_{i=1}^n \sigma_i I(t) dW_i.    
\end{equation*}
To further simplify, the following definitions are introduced, \begin{equation*}
    \sigma_I^2 = \frac{1}{n^2}\sum \limits_{i, j = 1}^n\rho_{ij}\sigma_i \sigma_j, \quad  q_I = \frac{1}{n}\sum \limits_{i=1}^n(q_i+\frac{1}{2}\sigma_i^2) - \frac{1}{2}\sigma_I^2, \quad W = \frac{1}{n\sigma_I}\sum \limits_{i=1}^n \sigma_i W_i.   
\end{equation*}
With these definitions, the dynamics of $I(t)$ reduce to\begin{equation}
    dI(t) = (r-q_I)I(t)dt + \sigma_II(t) dW. \label{99}
\end{equation}

Note that\begin{equation*}
dW dW = \frac{1}{n^2\sigma_I^2}\sum \limits_{i, j = 1}^n\sigma_i \sigma_j dW_i dW_j = 
 \frac{1}{n^2\sigma_I^2}\sum \limits_{i, j = 1}^n\sigma_i \sigma_j \rho_{ij} dt = dt   
\end{equation*}
By L$\acute{\text{e}}$vy’s Characterization Theorem of Brownian Motion \cite{shreve2004stochastic}, the quadratic variation of $W$ is $t$, then $W$ is a standard Brownian motion. Thus, from Eq.~(\ref{99}), $I_t$ is a geometric Brownian motion process. This enables us to use the binomial tree method to solve a one-dimensional option pricing problem with dividend $q_I$ and volatility $\sigma_I$ as an accurate approximation of the exact solution, and can be used as a benchmark to evaluate the accuracy of our methods. 

The method for constructing $g_2$ here is still to perform a first-order Taylor expansion on its European analytical solution. Take \begin{equation*}
    d_0 = -\frac{1}{\sigma_I \sqrt{\tau}}\big( \ln{\frac{I}{K}} + (r-q_I)\tau \big).   
\end{equation*}
Then the expression of $g_2$ is \begin{equation*}
    g_2(s_1, \cdots, s_n, t) = N(d_0)[Ke^{-r\tau}-Ie^{-q_I\tau}] + \frac{\sigma_I \sqrt{\tau}}{2\sqrt{2\pi}}e^{-\frac{d_0^2}{2}} \cdot (Ke^{-r\tau} + I)    
\end{equation*}
Such $g_2$ exactly satisfies the terminal condition. We take $g_1(s_1, \cdots, s_n, t) = T-t$ in our ETCNN.

The experiment considers cases with $n = 2, 3, 4, 5$ underlying assets, using the parameters $K=100, T=1, r = 0.05$. For the case of $n=5$ assets, we arbitrarily assign parameter values as follows, 
\begin{align*}
    d &= [0.02, 0.03, 0.04, 0.05, 0.03],    \notag \\ \sigma &= [0.15, 0.2, 0.25, 0.3, 0.22],    \notag \\ \rho &= \begin{pmatrix}
             1 & 0.2 & 0.3 & 0.1 & 0.4\\0.2 & 1 & 0.25 & 0.15 & 0.3 \\ 0.3 & 0.25 & 1 & 0.2 & 0.23 \\ 0.1 & 0.15  & 0.2& 1 & 0.26 \\ 0.4 & 0.3 & 0.23 & 0.26 & 1
         \end{pmatrix}.   
\end{align*}
Here, $d$ denotes the dividend rates. The dividend rate of $s_i$ is $d_i$, the $i$-th element of $d$. $\sigma$ denotes the volatility, where $\sigma_i$ is the volatility of $s_i$. The correlation matrix $\rho$ specifies the correlations between the assets, where $\rho_{ij}$ is the correlation between $s_i$ and $s_j$. For scenarios with $n = 2, 3, 4$, the dividend rates, volatilities, and correlation matrices are derived by taking the first $n$ elements of $d$ and $\sigma$, along with the $n\times n$ submatrix from the upper-left corner of $\rho$.
Input normalization is applied to both PINN and ETCNN. For $n = 2, 3$, the number of sampling points for $\mathcal{L}_{tc}$ are set to $N_{tc} = 8192$. For $n = 4, 5$, the number of sampling points is increased to $N_{tc} = 16384$ to accommodate the higher input dimensionality. For the other three loss terms, the sampling points are set to  $N_{bs} = N_{tv} = N_{eq} = 4N_{tc}$. 
Training is performed on the range $[0, 400]^n \times [0, T]$, while accuracy is evaluated on the interval $[80, 120]^n \times [0, T]$. The results of the experiments are summarized in Table \ref{table7}.

\begin{table}[H]
    \centering
\renewcommand{\arraystretch}{1.5} 
    \caption{Relative $L^2$ error and MAE for options on geometric average of $n$ assets. The input of networks in PINN and ETCNN has $n+1$ dimensions.  Both PINN and ETCNN apply the input normalization layer.}
    \begin{tabular}{cc|cccc}
    \hline
        Method & Error & $n=2$ & $n=3$ & $n=4$ & $n=5$ \\ \hline
        \multirow{2}{*}{PINN} & Rel. $L^2$ error & $2.00\times 10^{-2}$ & $2.79\times 10^{-2}$  & $1.48\times 10^{-1}$ & $1.72\times 10^{-1}$\\
        &MAE & $7.75\times 10^{-1}$  & $1.15\times 10^{0}$ & $2.53\times 10^{0}$ & $2.02\times 10^{0}$\\ \hline
        \multirow{2}{*}{ETCNN} & Rel. $L^2$ error & $1.53\times 10^{-3}$ & $1.69\times 10^{-3}$  & $1.07\times 10^{-3}$ & $1.24\times 10^{-3}$\\
        &MAE & $2.49\times 10^{-2}$  & $2.45\times 10^{-2}$ & $3.35\times 10^{-2}$ & $2.18\times 10^{-2}$\\ \hline
    \end{tabular}

    \label{table7}
\end{table}

The experimental results indicate that ETCNN significantly outperforms PINN in both relative $L^2$ error and MAE, achieving improvements of 1–2 orders of magnitude. This performance advantage becomes increasingly evident as the input dimensionality grows.
The performance of PINN deteriorates significantly as the dimensionality of the input space increases. When the number of underlying assets reaches 4 or more (i.e., when the input dimensionality exceeds 5), the error rates for PINN often exceed 
$10 ^ {-1}$, which is not applicable in practice. In contrast, ETCNN demonstrates remarkable robustness to the dimensionality increase, maintaining an accuracy level of approximately  $10 ^ {-3}$  even in high-dimensional scenarios. This stability and precision suggest that ETCNN is sufficiently accurate to be applied in real-world option pricing tasks.

\subsection{Call-on-Max American Options}
\label{Sec4.4}
In this example, we study call-on-max options. 
Let \begin{equation*}
\mathcal{F}(V(s_1, \cdots, s_n, t)) = \frac{\partial V}{\partial t} + \frac{1}{2} \sum \limits_{i,j=1}^n\sigma_i \sigma_j \rho_{ij}s_is_j \frac{\partial^2 V}{\partial s_i \partial s_j} + \sum \limits_{i=1}^n (r-q_i)s_i\frac{\partial V}{\partial s_i}-rV.       
\end{equation*}
The value of the American call option on the maximum value of $n$ underlying assets satisfies the following BSM equations,
\begin{numcases}{} \quad
V(s_1, \cdots, s_n, t) \geq (\max (s_1, \cdots, s_n)-K)^+,\quad  \forall s_i \geq 0,  t \in [0, T),    \notag \\ \quad
\mathcal{F}(V(s_1, \cdots, s_n, t)) \leq 0,\quad  \forall s_i \geq 0, t \in [0, T),    \notag \\ \quad \mathcal{F}(V(s_1, \cdots, s_n, t)) \cdot \big(V-(\max (s_1, \cdots, s_n)-K)^+\big) = 0, \quad  \forall s_i \geq 0, t \in [0, T), \quad     \notag \\ \quad V(s_1, \cdots, s_n, T) = (\max (s_1, \cdots, s_n)-K)^+, \quad \forall s_i \geq 0.   \notag 
\end{numcases}
This option is of research significance for two reasons. First, options on the maximum of two or more asset prices are widely used in practice. Examples include corporate bonds and managerial contracts with warrants \cite{broadie1997valuation, detemple2005american}. Second, the price of call-on-max options can be used to determine the prices of other related options, such as call-on-min, put-on-max, and put-on-min options \cite{stulz1982options}.

This section analyzes the American call option on the maximum of two assets, setting the strike price $K = 100$ and maturity $T=1$.
To demonstrate the universality and robustness of our method, we consider four market scenarios: low-volatility low-dividend, low-volatility high-dividend, high-volatility low-dividend, and high-volatility high-dividend. The dividend level here is relative to the risk-free rate $r$. For each scenario, we provide a representative example, with parameters specified as follows.

\begin{table}[H]
    \centering
\renewcommand{\arraystretch}{1.5} 
    \caption{Parameters for examples in each scenario.} 
    \begin{tabular}{cc|cccccc}
    \hline
        Number & Description & $\sigma_1$ & $\sigma_2$ & $\rho$ & $q_1$ & $q_2$ & $r$ \\
        \hline
        Scenario 1 & low-volatility low-dividend & $0.15$ & $0.25$  & $0.2$ & $0.02$& $0.04$& $0.06$\\ 
        Scenario 2&low-volatility high-dividend & $0.15$ & $0.25$  & $0.2$ & $0.03$& $0.06$& $0.02$\\ 
        Scenario 3 &   high-volatility low-dividend & $0.4$ & $0.3$  & $0.2$ & $0.02$& $0.04$& $0.06$\\  
        Scenario 4 &high-volatility high-dividend& $0.4$ & $0.3$  & $0.2$ & $0.03$& $0.06$& $0.02$\\ \hline
    \end{tabular}

    \label{table8}
\end{table}

\subsubsection{Exact Terminal Function Design}

This section explores the properties of the above differential equations and designs an exact terminal function $g_2$. It also illustrates the stopping region and provides the representation of the early exercise premium for call-on-max options \cite{broadie1997valuation, detemple2005american}.
Let the value of the option be $V = V(s_1, s_2, t)$. The payoff function is $\Phi(s_1, s_2) = (\max(s_1, s_2)-K)^+.$ The stopping region is the set \begin{equation*}
    \mathcal{S} = \{(s_1, s_2, t): V(s_1, s_2, t) = (\max(s_1, s_2) -K)^+\}     
\end{equation*} of price-date points on which the value of the options equals the immediate exercise payoff. The continuation region is defined as the complementary region of $\mathcal{S}$, i.e. \begin{equation*}
    \mathcal{C} = \{V(s_1, s_2, t):V(s_1, s_2, t) > (\max(s_1, s_2) -K)^+\}.   
\end{equation*}
Let $\mathcal{G}_i = \{(s_1, s_2, t): s_i = \max(s_1, s_2)\}$. The subregion $\mathcal{S}_i = \mathcal{S}\cap \mathcal{G}_i$ is the subset of $\mathcal{S}$ where asset $i$ is more expensive. The $t$-section, defined as $\mathcal{S}(t) = \{(s_1, s_2):(s_1, s_2, t) \in \mathcal{S}\}$ is the set of price pairs $(s_1, s_2)$ in the stopping region at the fixed time $t$. Similarly define $\mathcal{S}_i(t) = \{(s_1, s_2):(s_1, s_2, t) \in \mathcal{S}_i\}$. Based on these definitions, we can finally define the free boundary functions of multi-asset options, \begin{align*}
    S_1^*(s_2, t) = \inf \{s_1 \in \mathbb{R}^+: (s_1, s_2) \in \mathcal{S}_1(t)\},    \notag \\ S_2^*(s_1, t) = \inf \{s_2 \in \mathbb{R}^+: (s_1, s_2) \in \mathcal{S}_2(t)\}.    
\end{align*}
These two functions represent the boundary of the $t$-sections $\mathcal{S}_1(t)$ and $\mathcal{S}_2(t)$ respectively. Then the value of an American option on the maximum of two assets has the early exercise premium representation \begin{equation*}
    V^a(s_1, s_2, t) = V^e(s_1, s_2, t) + p(s_1, s_2, t),  
\end{equation*}
where the premium function $p(s_1, s_2, t)$ has two components, \begin{align*}
    p(s_1, s_2, t) &= p_1(s_1, s_2, t; S_1^*) + p_2(s_1, s_2, t; S_2^*),    \notag \\ p_1(s_1, s_2, t; S_1^*) &= \mathbf{E}_{t,s}\int_t^Te^{-r(u-t)}(d_1s_1(u)-rK)\textit{I}\{s_1\leq S_1^*(s_2, t)\}du,    \notag \\ p_2(s_1, s_2, t; S_2^*)&=\mathbf{E}_{t,s}\int_t^Te^{-r(u-t)}(d_2s_2(u)-rK)\textit{I}\{s_2\leq S_2^*(s_1, t)\}du.  
\end{align*}
$p_i(s_1, s_2, t; S_i^*)$ is defined for the continuous surface $S^*_i$. The free boundaries $S^*_1$ and $S^*_2$ are the solutions to the system of recursive integral equations, \begin{align*}
    S_1^*(s_2, t) - K &= V^a(S_1^*(s_2, t), s_2, t)  \\ &= V^e(S_1^*(s_2, t), s_2, t) + p_1(S_1^*(s_2, t), s_2, t; S_1^*) + p_2(S_1^*(s_2, t), s_2, t; S_2^*), \\ S_2^*(s_1, t) - K &= V^a(s_1, S_2^*(s_1, t), t) \\ &= V^e(s_1, S_2^*(s_1, t), t) + p_1(s_1, S_2^*(s_1, t), t; S_1^*) + p_2(s_1, S_2^*(s_1, t), t; S_2^*),
\end{align*}
Therefore, $V^a(s_1, s_2, t), S_1^*(s_2, t)$ and $S_2^*(s_1,t)$ form a coupled system of equations, which includes integration and expectation calculation. This makes it very difficult to solve $V^a(s_1, s_2, t)$ and find the free boundaries.

$V^e(s_1, s_2, t)$ is the value of the counterpart European option. Johnson \cite{johnson1987options} and Stulz \cite{stulz1982options} give the analytical formulas for the European call options on the maximum of two asset prices without dividends. Here, we derive the analytical formula for European call-on-max options with dividends. 
Consider the following notations,\begin{align*}
    d_1(s_i, \sigma; K, \tau) &= \frac{1}{\sigma \sqrt{\tau}}\Big[\ln(\frac{s_i}{K}) + (r-q_i+\frac{1}{2}\sigma^2)\tau)\Big], \quad i = 1, 2, \\ d_2(s_i, \sigma; K, \tau) &= \frac{1}{\sigma \sqrt{\tau}}\Big[\ln(\frac{s_i}{K}) + (r-q_i-\frac{1}{2}\sigma^2)\tau)\Big],\quad  i = 1, 2, \\ d_1'(s_i, s_j, \sigma; \tau) &= \frac{1}{\sigma \sqrt{\tau}}\Big[\ln(\frac{s_i}{s_j}) + (q_j-q_i+\frac{1}{2}\sigma^2)\tau)\Big],\quad  i, j = 1, 2,\\ d_2'(s_i, s_j, \sigma; \tau) &= \frac{1}{\sigma \sqrt{\tau}}\Big[\ln(\frac{s_i}{s_j}) + (q_j-q_i-\frac{1}{2}\sigma^2)\tau)\Big], \quad i, j = 1, 2.
\end{align*}
The parameters involved in this derivation include \begin{align*}
    \sigma_{12}^2 = \sigma_1^2-2\rho\sigma_1\sigma_2+\sigma_2^2, \quad   \rho_1 = \frac{\sigma_1-\rho \sigma_2}{\sigma_{12}},\quad \rho_2 = \frac{\sigma_2-\rho \sigma_1}{\sigma_{12}}.   
\end{align*}
With $\tau = T-t$ denoting the time to maturity, the analytical formula for European call-on-max options with dividends is as follows, \begin{align*}
    V^e(s_1, s_2, t)& = s_1e^{-q_1\tau}N_2(d_1(s_1, \sigma_1; K, \tau), d_1'(s_1, s_2, \sigma_{12};\tau), \rho_1) \notag \\ &+ s_2e^{-q_2\tau}N_2(d_1(s_2, \sigma_2; K, \tau), d_1'(s_2, s_1, \sigma_{12};\tau), \rho_2) \notag  \\ &-Ke^{-r\tau}\big[1-N_2(-d_2(s_1, \sigma_1; K, \tau), -d_2(s_2, \sigma_2; K, \tau), \rho)\big].    
\end{align*}
Here $N_2(x, y, \rho)$ is the cumulative distribution function of the bivariate standard normal distribution. Note that $V^e(s_1, s_2, t)$ satisfies the same terminal conditions as $V^a(s_1, s_2, t)$. Since it has a complex form, it is difficult to find a $d_0$ like in the single-asset case. Therefore we choose $g_2(s_1, s_2, t) = V^e(s_1, s_2, t)$ in this experiment. The definition of $g_1$ is still $g_1(s_1, s_2, t) = T-t$.

Calculating $V^e(s_1, s_2, t)$ is time-expensive due to the double integrals involved in the bivariate normal distribution. Moreover, when calculating the loss function, its derivatives are involved, which are more complicated.
To accelerate the training process, we implement the following two techniques.
First, we employ the method proposed in \cite{Drezner1990, genz2004numerical} to numerically approximate the bivariate normal distribution. Gauss-Legendre integration rule and Taylor expansion approximation are used here.
The error caused by using the approximation here is greatly smaller than the errors elsewhere. Therefore, it would not affect the accuracy.
Second, given that $\mathcal{F}(V^a)\leq 0, \mathcal{F}(V^e)= 0$, if the network prediction $g_1u_{NN}+g_2$ represents the solution of $V^a$ and $g_2 = V^e$, then it follows that $\mathcal{F}(g_1u_{NN})\leq 0$. Therefore in the loss term $\mathcal{L}_{bs}$, we use $\mathcal{F}(g_1u_{NN})$ instead of using $\mathcal{F}(g_1u_{NN} + g2)$, as $\mathcal{F}(g_2)= 0$ is automatically satisfied. This approach removes the need to differentiate $g_2$, thereby saving computation time.

\begin{figure}[!t]
    \centering  \includegraphics[width=0.82\textwidth]{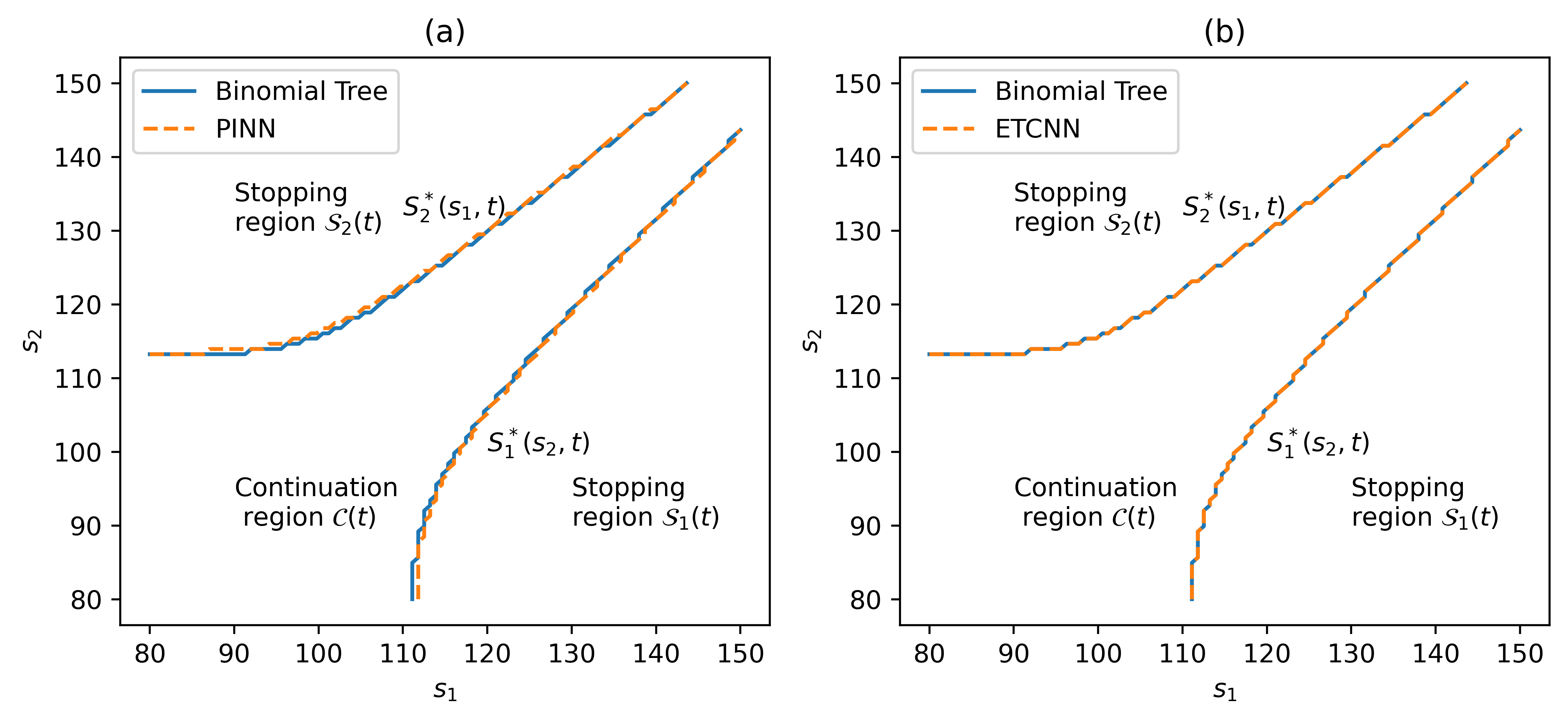}\\
    \caption{Illustration of stopping regions and continuation region for a call-on-max option with $\sigma_1 = 0.1, \sigma_2 = 0.25, \rho = 0.2, d_1 = 0.02, d_2 = 0.04, r = 0.06, K = 100, T = 1$, at time $t=0.75$ (i.e., this option is three months before maturity). $(a)$, The free boundaries $S_1^*(s_2, t)$ and $S_2^*(s_1, t)$ obtained by BT and PINN. $(b)$, The free boundaries $S_1^*(s_2, t)$ and $S_2^*(s_1, t)$ obtained by BT and ETCNN.}
        \label{fig10}
\end{figure}

\subsubsection{Numerical Results}

The solution obtained by BT with $N = 400$ serves as the reference solution. Here $N$ is reduced by an order of magnitude compared with the previous cases because the complexity increases exponentially with $N$ in high-dimensional situations. 
The network input consists of three dimensions, $s_1, s_2$, and $t$, with normalization applied to $s_1$ and $s_2$. For the loss calculations, $N_{tc} = 8192$ sampling points are used to compute 
$\mathcal{L}_{tc}$, while $N_{bs} = N_{tv} = N_{eq} = 4N_{tc}$ are used for the other three loss terms. The network is trained over the domain $[0, 400]^2 \times [0, T]$ with error evaluation performed on $[80, 150]^2 \times [0, T]$. Both PINN and ETCNN are evaluated under four typical market scenarios, with the results summarized in Table \ref{table9}.

\begin{table}[!h]
    \centering
\renewcommand{\arraystretch}{1.5} 
    \caption{Relative $L^2$ error and MAE for the four market scenarios. The input of networks for PINN and ETCNN has 3 dimensions. Both PINN and ETCNN implement the input normalization.}
    \begin{tabular}{cc|cccc}
    \hline
        Method & Error  & Scenario 1& Scenario 2 & Scenario 3 & Scenario 4 \\ \hline
        \multirow{2}{*}{PINN} & Rel. $L^2$ error & $3.18\times 10^{-2}$ & $3.30\times 10^{-2}$  & $3.55\times 10^{-2}$ & $4.53\times 10^{-2}$\\ 
        &MAE & $1.53\times 10^{0}$  & $1.72\times 10^{0}$ & $2.25\times 10^{0}$ & $2.91\times 10^{0}$\\ \hline
        \multirow{2}{*}{ETCNN} & Rel. $L^2$ error & $1.13\times 10^{-4}$ & $2.01\times 10^{-4}$  & $9.84\times 10^{-5}$ & $2.12\times 10^{-4}$\\ 
        &MAE & $1.77\times 10^{-2}$  & $3.53\times 10^{-2}$ & $2.06\times 10^{-2}$ & $5.06\times 10^{-2}$\\ \hline
    \end{tabular}

    \label{table9}
\end{table}

As demonstrated in Table \ref{table9}, our approach has a significant improvement over PINN, reducing the $L^2$ error by two orders of magnitude and achieving an MAE on the order of $10^{-2}$. In high-dimensional cases, traditional numerical methods are hard to implement, while PINN is difficult to calculate accurate solutions. However, our ECTNN is easy to implement and can achieve high accuracy. Furthermore, the four market scenarios presented cover a broad range of market conditions, and our method performs well across all scenarios, indicating its robustness and versatility.

Figure \ref{fig10} illustrates the free boundaries obtained by PINN and ETCNN for Scenario 1 at $t=0.75$. Since the option has a maturity of $T = 1$, $t=0.75$ means it is $0.25$ year, or three months before maturity. The free boundaries computed from the reference solution are considered as the true free boundaries. 
As shown in it, the free boundaries identified by our ETCNN method align closely with the true ones, while the free boundaries obtained by PINN exhibit obvious deviations, particularly at lower asset prices. This comparison indicates that our method achieves relatively high accuracy in determining complex free boundaries in high-dimensional problems.

\section{Conclusions}
\label{sec5}
This study introduces the exact terminal condition neural network (ETCNN) framework to solve the Black-Scholes-Merton equation with inequality constraints.
By incorporating exact terminal functions that exactly satisfy terminal conditions and capture the singular behaviors of the true solution, ETCNN effectively reduces approximation complexity and improves accuracy compared to other neural network-based methods.
The proposed approach has been tested across a variety of scenarios, including both single-asset and multi-asset cases with different parameters. 
Numerical results further demonstrate that ETCNN consistently outperforms both traditional approaches and other neural network methods in terms of accuracy, while maintaining computational efficiency. 

Despite these promising results, several directions for future research remain. 
Future studies could explore extending this approach to options with more complex terminal conditions, such as Asian options, barrier options, and other exotic derivatives. 
Moreover, the framework could be adapted to more advanced models, such as local volatility models and stochastic volatility models, which present additional mathematical and computational challenges.
Addressing these challenges could significantly expand the scope and applicability of the ETCNN framework in financial modeling.

\section*{Acknowledgments}
The authors sincerely appreciate the anonymous reviewers for their valuable comments and suggestions, which significantly enhance the quality and clarity of this work. 
The work of Lu and Zhang was funded by the Strategic Priority Research Program of Chinese Academy of Sciences (Grant No. XDB0500000)
and the National Natural Science Foundation of China (Grant No. 12371413, No. 22073110).
The work of Guo was supported by the National Natural Science Foundation of China (Grant No. 12371438).  
The AI-driven experiments, simulations and model training were performed on the GPU computing platform of the Academy of Mathematics and Systems Science, Chinese Academy of Sciences
and the robotic AI-Scientist platform of Chinese Academy of Sciences.





\bibliographystyle{elsarticle-num}
\bibliography{references}

\end{document}